\documentclass[journal]{IEEEtran}
\usepackage{cite}

\ifCLASSINFOpdf
\usepackage[pdftex]{graphicx}
\else
 \usepackage[dvips]{graphicx}
\fi
\usepackage{amsmath}
\usepackage{amssymb}
\usepackage{bm}
\usepackage{array}

\usepackage{xcolor}

\usepackage{multirow}

\begin{document}
%
\title{Mixed-gradients Distributed Filtered Reference Least Mean Square Algorithm -- A Robust Distributed Multichannel Active Noise Control Algorithm}
%
%
%

\author{Junwei Ji,~\IEEEmembership{Student Member,~IEEE,}
        Dongyuan Shi,~\IEEEmembership{Senior Member,~IEEE,}
        and Woon-Seng Gan,~\IEEEmembership{Senior Member,~IEEE}
\thanks{Junwei Ji and Woon-Seng Gan are with the School of Electrical and Electronic Engineering, Nanyang Technological University, Singapore 639798, Singapore (e-mail: JUNWEI002@e.ntu.edu.sg; ewsgan@ntu.edu.sg).}
 \thanks{Dongyuan Shi is with the Center of Intelligent Acoustics and Immersive Communications, Northwestern Polytechnical University, Xi'an, China 710071 (e-mail: dongyuan.shi@nwpu.edu.cn).}
}
\maketitle

\begin{abstract}

Distributed multichannel active noise control (DMCANC), which utilizes multiple individual processors to achieve a global noise reduction performance comparable to conventional centralized multichannel active noise control (MCANC), has become increasingly attractive due to its high computational efficiency. However, the majority of current DMCANC algorithms disregard the impact of crosstalk across nodes and impose the assumption of an ideal network devoid of communication limitations, which is an unrealistic assumption. Therefore, this work presents a robust DMCANC algorithm that employs the compensating filter to mitigate the impact of crosstalk. The proposed solution enhances the DMCANC system's flexibility and security by utilizing local gradients instead of local control filters to convey enhanced information, resulting in a mixed-gradients distributed filtered reference least mean square (MGDFxLMS) algorithm. The performance investigation demonstrates that the proposed approach performs well with the centralized method. Furthermore, to address the issue of communication delay in the distributed network, a practical strategy that auto-shrinks the step size value in response to the delayed samples is implemented to improve the system's resilience. The numerical simulation results demonstrate the efficacy of the proposed auto-shrink step size MGDFxLMS (ASSS-MGDFxLMS) algorithm across various communication delays, highlighting its practical value.

\end{abstract}


%
\IEEEpeerreviewmaketitle

\section{Introduction}
%
%
%
%
\IEEEPARstart{B}{eing} exposed to noise for a long time not only affects our physical health but can also cause psychological disorders \cite{peris2020noise}. Passive noise control which uses barriers to block the propagation of noise can reduce high-frequency components \cite{Kuo1999ANC}. Some persistent low-frequency noise can be effectively suppressed by an active noise control (ANC) system, which generates an anti-noise that has the same amplitude but the opposite phase as the primary noise to eliminate unwanted sound \cite{lueg1936process}. To adapt to the various noise, the filtered reference least mean square (FxLMS) \cite{Morgan1980FXLMS} algorithm is widely used in an ANC field. Although ANC technology has achieved commercial success, some practical issues still remain \cite{Lam2021Ten,Kajikaea2012ANC}, such as output saturation \cite{Shi2019TwoGDFXLMS,Shi2021OLFXLMS,Lai2023MOVMFXLMS}, limited sensors placement \cite{moreau2008review,liu2024relative,sun2022spatial,shi2019selective}, low signal-to-noise ratio reference signal \cite{Shen2022Wireless,Shen2024wireless} and slow adaption to dynamic noise \cite{Shi2023Transferable,Luo2023GFANC,Luo2024ImplemSFANC}.

A multichannel active noise control (MCANC) system, which deploys multiple secondary sources and multiple error sensors to obtain a global noise reduction over a large region has been of great interest in quieting the environment recently \cite{iwai2019multichannel,lorente2014gpu}. The Conventional MCANC system, which is also known as the centralized noise control, employs a single controller to process all inputs and outputs and to update the system simultaneously by gathering all the error signals. However, its massive computational burden places a high demand on the controller's performance and resources, thus leading to high implementation costs. To reduce the computational complexity of centralized control strategy, some efficient algorithms have also been proposed, such as the partial update algorithm \cite{shi2018partial}, mixed error approach \cite{murao2017mixed}, and the block coordinate descent based algorithm \cite{shi2021block}. Another approach is the decentralized control strategy \cite{Zhang2019Decentralized,george2012particle}, which allocates the computational tasks into several independent controllers, where each controller only considers its error signal to update the control filter individually, resulting in local noise cancellation. However, neglecting the mutual acoustic crosstalk effect increases the risk of instability in decentralized algorithms \cite{Kuo1999ANC}. 

For the sake of balancing the advantages of centralized and decentralized control strategy, the distributed MCANC (DMCANC) system is developed \cite{Ferrer2015DistributedANC,Ferrer2017Distributed,Antonanzas2016Blockwise,Antonanzas2017IncrementalANC}. It has several ANC nodes, each of which may include one or more secondary sources, one or more error sensors, and an ANC controller for signal processing and communication. In DMCANC, each node processes its own received signal independently, while the nodes exchange certain information with each other to ensure global noise reduction performance. Additionally, wireless network transmission allows sensors and secondary sources to operate independently of the central processor's location, enabling more flexible sensor placement. Distributed strategy allows the ANC nodes to be more flexible in arranging the desired noise reduction area compared to the centralized strategy. 

Inspired by \cite{Lopes2007Incremental,Lopes2008DiffusionLMS,Aravinthan2011Wireless}, some control strategies are available for DMCANC, such as incremental strategy and diffusion strategy. Incremental strategy \cite{Ferrer2015DistributedANC,Lorente2015DistributedANC,Ferrer2021Affine,Antonanzas2016AffineDistributed} updates node by node requiring high communication requirements, while diffusion strategy \cite{Antonanzas2015Diffusion,Song2016DiffusionANC,Kukde2019DiffusionANC,Chen2022DistributedANC} exchanges information locally and cooperate only with their neighbours, without the need for sharing or requiring any global information. Hence, the latter is widely considered in the conventional DMCANC system. Conventional diffusion filtered reference least mean square (DFxLMS) algorithm utilizes topology-based combination rules to integrate received data \cite{Chu2019DiffusionANC,Chu2020DiffusionANC}, which is regarded as spatial smoothing \cite{Chu2021Combination}, leading to an ineffective control on asymmetric path. To solve this issue, an augmented DFxLMS (ADFxLMS) algorithm \cite{Li2023Distributed} is proposed at the cost of communication load. Following this work, Li et.al. describe a bidirectional communication method to improve ADFxLMS algorithm's communication burden \cite{Li2023AugmentedDiffusion}. However, as far as we are aware, the majority of DMCANC systems ignore the crosstalk effect \cite{Ji2023Distributed} between nodes and also assume that each node receives prompt information from other nodes within an iteration, which does not consider communication restrictions and latency \cite{Lee2015Distributed}. These assumptions are not practical and will lead to poor noise control performance.

Therefore, a robust DMCANC algorithm is proposed, where the compensation filters are introduced to compensate for the difference in secondary paths between the nodes, resulting in a reduced cross-talk effect. Furthermore, a mixed-gradients distributed FxLMS (MGDFxLMS) algorithm is thus derived where the local gradient instead of the local control filter of each node is shared in the distributed network and then is aggregated by the compensation filters to update the global control filters, exhibiting a flexible and secure system. The performance analysis demonstrates that the proposed algorithm performs similarly to the conventional centralized algorithm. In addition, the analysis also reveals the effect of communication delays on the convergence of the algorithm. To counteract the instability caused by communication delays, a practical strategy is applied, where the step size value for each node is auto shrunk according to the delayed samples, leading to a more robust DMCANC system for the unstable distributed network. Compared with the existing DMCANC algorithm, the proposed auto-shrink step size MGDFxLMS (ASSS-MGDFxLMS) algorithm shows great practical significance.

The remainder of the paper is organized as follows: Section~\ref{sec:MANC} briefly introduces different control strategies of the MCANC system. Section~\ref{sec:method} describes the proposed DMCANC method as well as the auto-shrink step size strategy to overcome the issue of communication delay. In Section~\ref{sec:analysis}, the algorithm performance will be analysed to illustrate the effect of communication delay. Simulation results exhibited in Section~\ref{sec:sim} demonstrate the effectiveness of the proposed ASSS-MGDFxLMS algorithm under both an ideal and an unstable network. Finally, the conclusion is drawn in Section~\ref{sec:conclusion}.

\section{Multichannel ANC System}\label{sec:MANC}

\begin{figure}[!t]
    \centering
    \includegraphics[width = 0.9\columnwidth]{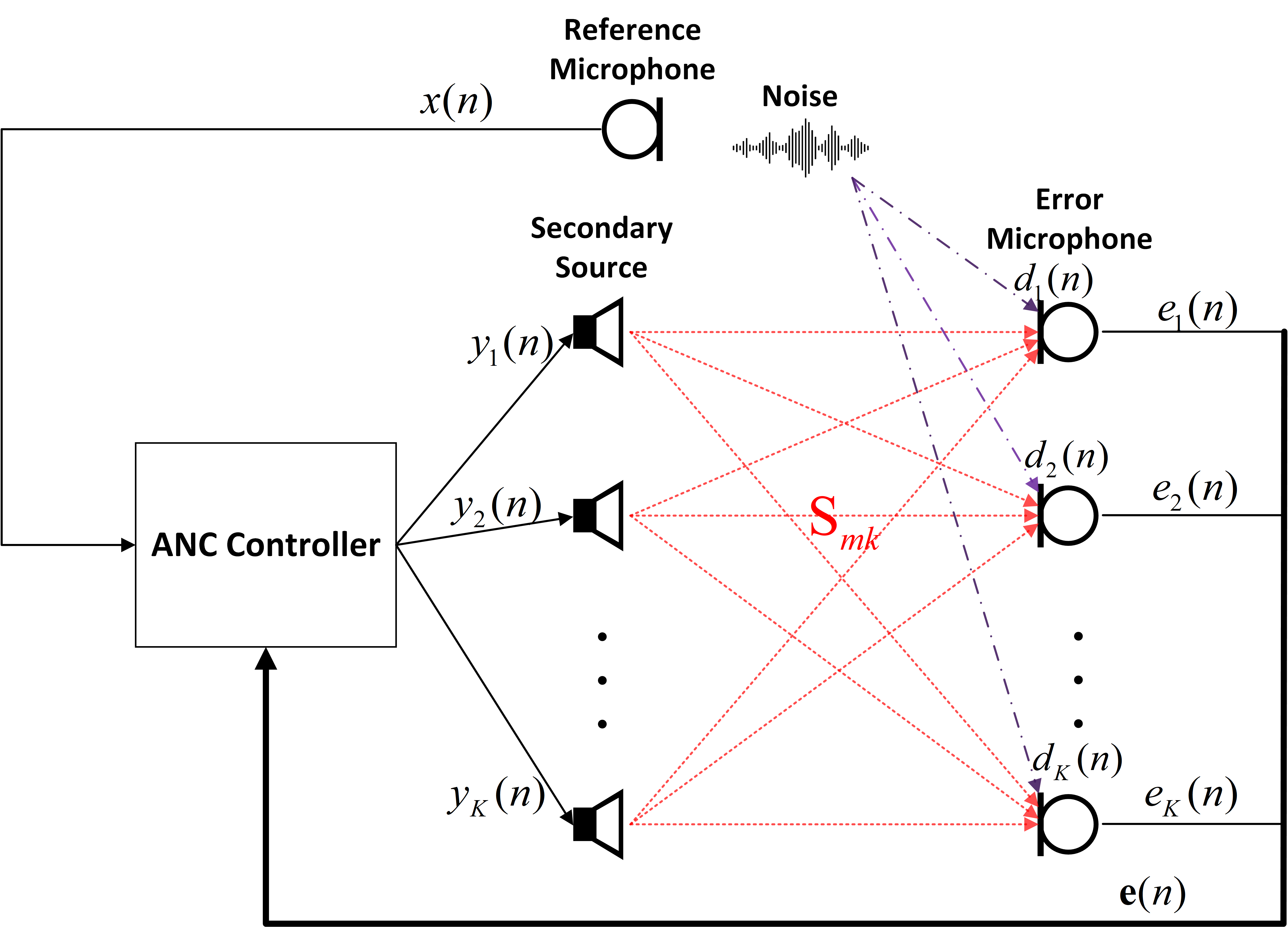}
    \caption{The schematic diagram of conventional multichannel ANC, where $S_{mk}$ represents the secondary path from the $k$th secondary source to the $m$th error sensor.}
    \label{fig1:MEANC}
\end{figure}

The multichannel ANC (MCANC) system is widely used to achieve a large quiet zone. One of the typical MCANC systems consisting of one reference sensor, $K$ secondary source, and $K$ error sensors is shown in Fig.~\ref{fig1:MEANC}, where the reference sensor captures the primary noise resulting in the reference signal $x(n)$, and it will be fed into the ANC controller to generate the control signals for each secondary source. Hence, the $k$th control signal is obtained from 
\begin{equation}\label{eq1:controlsignal}
    y_k(n) = \mathbf{w}_k^\mathrm{T}(n)\mathbf{x}(n), \quad k = 1,2,...,K,
\end{equation}
where $\mathbf{w}_k(n)=[w_{k,0}(n) \, w_{k,1}(n) \, \cdots \, w_{k,N-1}(n)]^\mathrm{T}$ and $\mathbf{x}(n)=[x(n) \, x(n-1) \, \cdots \, x(n-N+1)]^\mathrm{T}$ denote the $k$th control filter and reference vectors with the length of $N$, and $n$ is the time index. These control signals pass through secondary paths $S_{mk}$ resulting in anti-noise to suppress the disturbance $d_m(n)$. Here, the secondary path is regarded as the acoustic path from the secondary source to the error sensor. Hence, the residual error signal $e_m(n)$ measured by the $m$th error sensor is expressed as
\begin{equation}\label{eq2:error}
    e_{m}(n) = d_{m}(n) - \sum_{k=1}^{K}{y}_{k}(n)*{s}_{mk}(n), \quad m = 1,2,...,K,
\end{equation}
where $*$ denotes the linear convolution, and ${s}_{mk}(n)$ refers to the impulse response of secondary path $S_{mk}$ from the $k$th secondary source to the $m$th error sensor. 

To derive the above control filters and realize active noise control, there are different control strategies: centralized, decentralized, and distributed methods.

\subsection{Centralized MCANC}\label{subsec:cmanc}
The centralized control strategy is a widely used method that employs a single processor to handle all inputs, such as reference and error signals while generating the control signals. The block diagram of this strategy is shown in Fig.~\ref{fig1:MEANC}. Its cost function is defined as
\begin{equation}\label{eq3:costfunc}
    J = \sum_{m=1}^{K}\mathbb{E}[{e}^2_{m}(n)],
\end{equation}
where $\mathbb{E}[\cdot]$ represents the expectation operation. Based on the gradient descent method, its negative instantaneous gradients are used to update the control filter as
\begin{equation}\label{eq4:centralizedupdate}
    \mathbf{w}_k(n+1) = \mathbf{w}_k(n) + \mu \sum_{m=1}^K\mathbf{x}_{km}'(n)e_m(n),
\end{equation}
where $\mathbf{x}_{km}'(n)$ represents the filtered reference signal vector  obtained from
\begin{equation}\label{eq5:filteredx}
    \mathbf{x}_{km}'(n) = \hat{s}_{mk}(n)*\mathbf{x}(n).
\end{equation}
In above equation, $\hat{s}_{mk}(n)$ represents the estimate of the secondary path $s_{mk}(n)$ with $L$ taps. This equation is a well-known multichannel  FxLMS (MCFxLMS) algorithm \cite{Elliott1987MEANC}. However, its huge computational cost places high demands on the processor's performance, which makes it more difficult to implement in practice.

\subsection{Decentralized MCANC}
The decentralized technique separates the centralized MCANC system into multiple individual ANC units. Each ANC unit consists of a secondary source, an error sensor, and an ANC controller. Meanwhile, its cost function becomes
\begin{equation}\label{eq6:Decostfunc}
    J_D = \mathbb{E}[{e}^2_{m}(n)],
\end{equation}
and the updating equation of the control filter in $k$th controller is given by
\begin{equation}\label{eq7:decentralizedupdate}
    \mathbf{w}_k(n+1) = \mathbf{w}_k(n) + \mu_k \mathbf{x}_{kk}'(n)e_k(n),
\end{equation}
where $\mathbf{x}_{kk}'(n)$ is obtained from \eqref{eq5:filteredx} when $m=k$,  and $\mu_k$ denotes the step size. 
The equation~\eqref{eq7:decentralizedupdate} implicates that in decentralized MCANC, each channel only updates its control filter using its own error signal, leading to local noise control. Although the independent control technique significantly decreases the computing burden compared to the centralized strategy, it also leads to a degradation in noise reduction efficacy and the risk of instability due to cross-talk effects from other channels. 

\subsection{Distributed MCANC}
\begin{figure}[!t]
    \centering
    \includegraphics[width = 0.9\columnwidth]{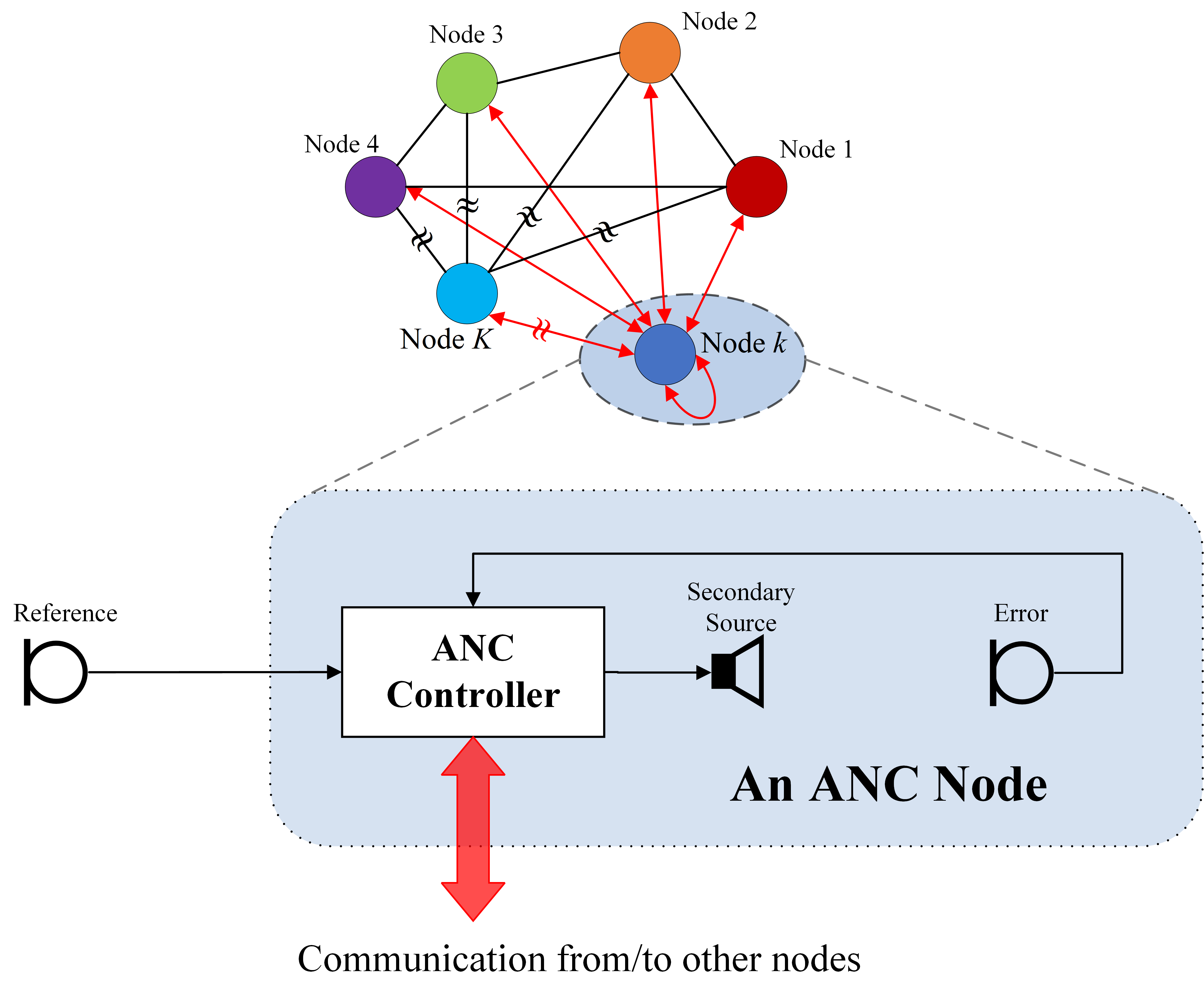}
    \caption{A DMCANC network, where each ANC node consists of a secondary source, an error sensor, and an ANC controller.}
    \label{fig3:ANCnode}
\end{figure}


The concept of distributed MCANC (DMCANC) is derived from a decentralized approach that involves distributing large computing tasks among multiple ANC nodes. Each node in this system functions as a basic ANC unit with a single controller, a single secondary source, and a single error microphone. In order to address the problem of instability in the decentralized MCANC, the ANC controller in each node performs signal processing and exchanges essential information, such as the control filter, with other nodes, as shown in Fig.~\ref{fig3:ANCnode}. Thus, conventional DMCANC consists of two primary processes: adaptation and combination.

During the adaptation phase, each node updates its individual local control filter, $\boldsymbol{\psi}_{k}(n)$, using its associated error signal as
\begin{equation}\label{eq8:adaptation}
    \boldsymbol{\psi}_{k}(n) = \mathbf{w}_k(n-1) + \mu_{k} \mathbf{x}_{kk}'(n)e_k(n).
\end{equation}
The global control filter for each node, $\mathbf{w}_k(n)$, is obtained from combining the local control filter with the received local control filters from other nodes, defined as
\begin{equation}\label{eq9:combination}
\mathbf{w}_{k}(n)=\sum_{l\in \mathcal{N}_{k}}a_{lk}\boldsymbol{\psi}_{l}(n),
\end{equation}
where $\mathcal{N}_k$ represents the neighbourhood of node $k$, and $a_{lk}$ stands for the combination weights, which satisfy $\Sigma_{l \in \mathcal{N}_k }a_{lk} = 1$. 
It is worth noting that the above processing is such that the adaption occurs before the combination, which is considered to be the Adapt-Then-Combine (ATC) technique \cite{Cattivelli2010Diffusion,Chu2020DiffusionANC,Li2023Distributed}. Conversely, swapping \eqref{eq8:adaptation} and \eqref{eq9:combination} so that the process of combining is carried out prior to adaptation is referred to as the Combine-Then-Adapt (CTA) approach \cite{Cattivelli2010Diffusion,Chu2019DiffusionANC}.


However, the topology-based combination rule is just a weighted sum of local control filters, resulting in a spatial average \cite{Chu2021Combination}. In practical applications, this technology requires the system's acoustic path to be symmetrical, which is unrealistic. Additionally, the acoustic crosstalk effect between nodes remains unsolved. Moreover, communication also holds a significant position in DMCANC. As a result of communication limitations, specific nodes cannot receive timely and valuable information, leading to instability in the system.

\begin{figure}[!t]
    \centering
    \includegraphics[width = 0.9\columnwidth]{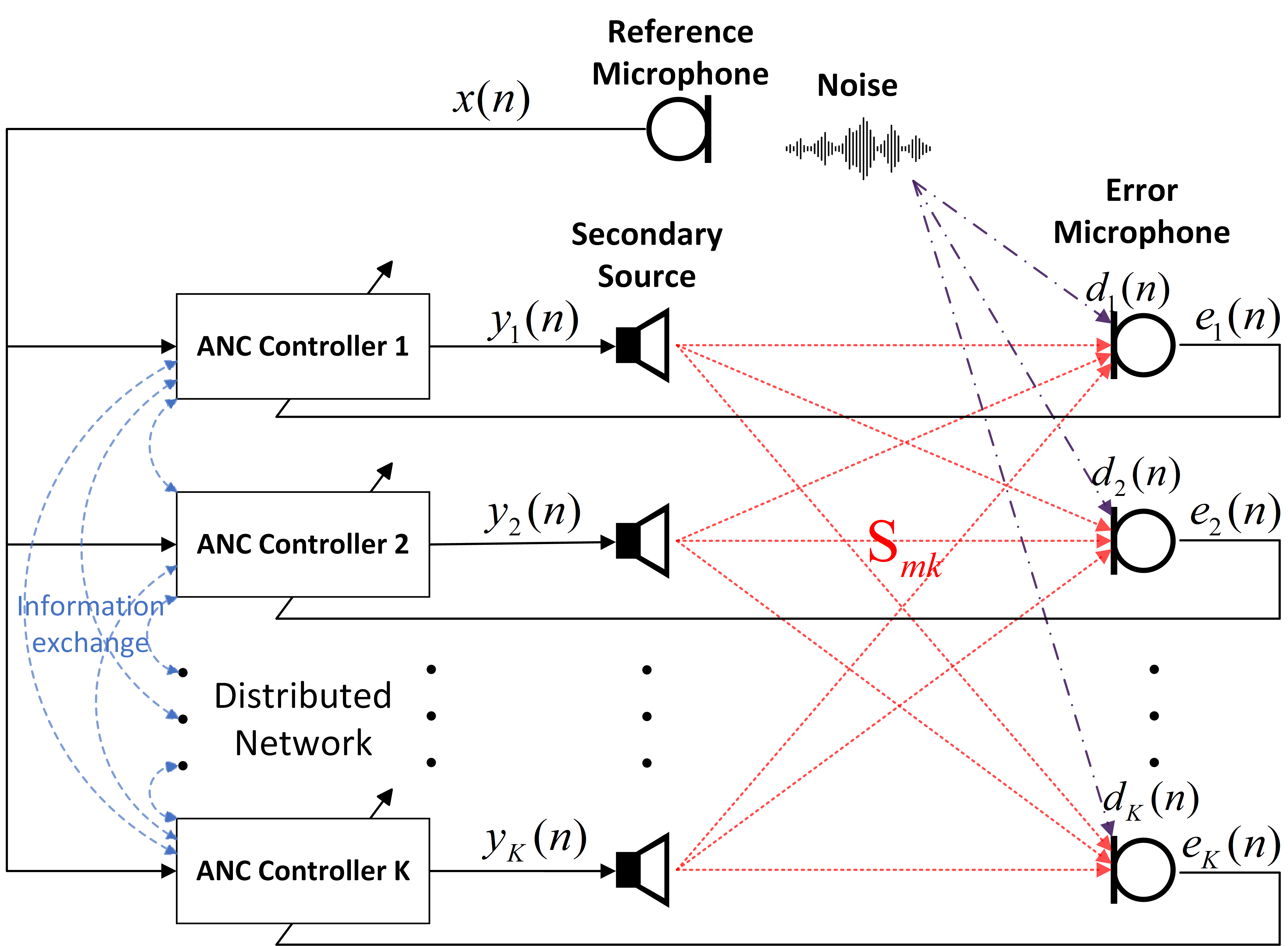}
    \caption{The schematic diagram of DMCANC, where each ANC controller exchanges information through a distributed network. The anti-noise wave generated by each ANC controller is transmitted to all error microphones, leading to inter-node cross-talk effects.}
    \label{fig2:DMANC}
\end{figure}

\section{Proposed Methodology}\label{sec:method}
This section presents a novel distributed MCANC algorithm derived from the conventional centralized MCFxLMS algorithm. In the proposed method, the compensation filters are introduced to combine with received local gradients to update the global control filters, resulting in a mixed-gradients distributed FxLMS (MGDFxLMS) algorithm. Furthermore, the auto-shrink step size strategy is applied to improve the system's robustness to communication delays.

\subsection{Compensation filters}
DMCANC consists of many ANC nodes, as illustrated in Fig.~\ref{fig3:ANCnode}. Each node is equipped with its own secondary path, referred to as the self-secondary path.
Since these ANC nodes may be close to each other, the current node may receive control signals generated from other nodes, as illustrated in Fig.~\ref{fig2:DMANC}, resulting in a cross-talk effect. The acoustic paths these signals pass through can be called cross-secondary paths. In general, the cross-secondary paths are longer than the self-secondary paths. 
Hence, the compensation filters, $c_{mk}(n)$, are introduced to make up for their difference \cite{Ji2023Distributed} described as
\begin{equation}\label{eq10:compensate}
 s_{mk}(n) = s_{mm}(n) * c_{mk}(n), \; (m\neq k),
\end{equation}
where $s_{mm}(n)$ and $s_{mk}(n)$ represent self-secondary path and cross-secondary path, respectively.

\begin{figure}[!t]
    \centering
    \includegraphics[width = 0.9\columnwidth]{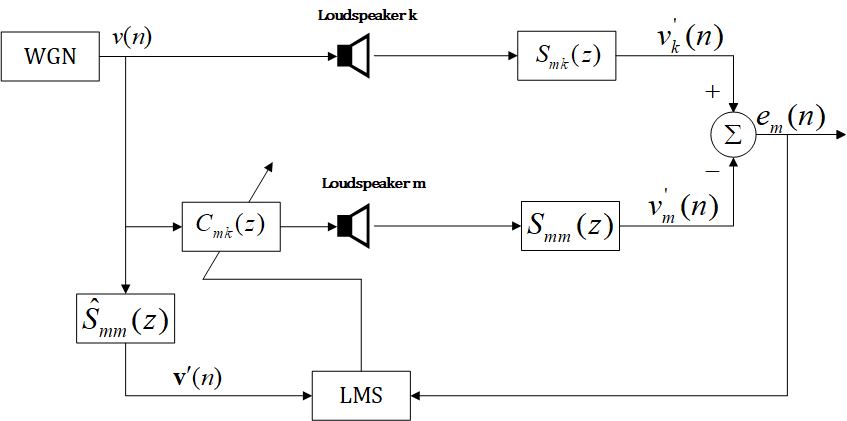}
    \caption{Block diagram of obtaining compensation filters using the FxLMS algorithm, where $S_{mk}(z)$ and $S_{mm}(z)$ are the cross and self secondary path respectively. $\hat{S}_{mm}(z)$ denotes the estimated self secondary path and $C_{mk}(z)$ represents the compensation filter.}
    \label{fig4:compensationfilter}
\end{figure}

These compensation filters can be obtained by using the FxLMS algorithm, whose block diagram is illustrated in Fig.~\ref{fig4:compensationfilter}. A white Gaussian noise (WGN), $v(n)$, is used to model the compensation filter, and its impulse response can be derived recursively as
\begin{equation}\label{eq11:compensationfilter}
    \mathbf{c}_{mk}(n+1) = \mathbf{c}_{mk}(n) + \mu_{c}\mathbf{{v}'}(n)e_m(n), (m\neq k),
\end{equation}
where $\mu_{c}$ represents the step size, and $\mathbf{{v}'}(n)$ denotes WGN vector $\mathbf{v}(n)$ filtered by estimated self secondary path expressed as:
\begin{equation}\label{eq:filteredwgn}
    \mathbf{{v}'}(n) = \hat{s}_{mm}(n) * \mathbf{v}(n).
\end{equation}
 The error signal is given by 
\begin{equation}\label{eq12:compenerror}
    e_{m}(n) = v'_{k}(n)-v'_{m}(n), (m \neq k),
\end{equation}
where $v'_{k}(n)$ is regarded as the desired signal that is captured by WGN through cross-secondary path and $v'_{m}(n)$ is formed by WGN through compensation filter and self-secondary path, which is defined as:
\begin{equation}\label{eq:desiredandanti}
    \begin{cases}
        {v}'_k(n) = & v(n) * s_{mk}(n), \\
        {v}'_m(n) = & \mathbf{c}_{mk}^{\mathrm{T}}(n)\mathbf{v}(n) * s_{mm}(n).
    \end{cases}
\end{equation}
It is worth noting that there are total $K(K-1)$ compensation filters that can be estimated node by node through offline training before operating ANC.

\subsection{Mixed gradients updating technique}
DMCANC allows each node to update independently according to its own error signals while ensuring noise reduction performance through the exchange of information between nodes. Considering the $k$th node, its error signal can be calculated as
\begin{equation}\label{eq13:nodekerr}
    e_k(n) = d_k(n) - y'_k(n) - \gamma_k(n),
\end{equation}
where $y'_k(n)$ denotes the anti-noise expressed as
\begin{equation}\label{eq14:antinoise}
     y'_k(n) = y_k(n) * s_{mk}(n),
\end{equation}
and $\gamma_k(n)$ is regarded as the interference caused by the other nodes given by
\begin{equation}\label{eq15:interference}
     \gamma_k(n) = \sum_{m=1,m\neq k }^K y_m(n)*s_{mk}(n).
\end{equation}

To achieve the noise reduction performance for the $k$th node, the local instantaneous cost function is defined as
\begin{equation}\label{eq16:localcostfunc}
     J_k(n) = e^2_k(n).
\end{equation}
Furthermore, the notation $\mathbf{w}_k(n)$ is defined as the global control filter. 
By taking the partial differential of \eqref{eq16:localcostfunc} with respect to $\mathbf{w}_k(n)$, the local gradient for the $k$th node can be derived as
\begin{equation}\label{eq17:localgradient}
    \boldsymbol{\nabla}_k(n) = -2  [\mathbf{x}(n)*\hat{s}_{kk}(n)]\cdot e_k(n).
\end{equation}

Nevertheless, the control signal produced by the $k$th node will also be sent to other nodes, resulting in cross-talk effects. Therefore, the error signal of the other node can be represented as
\begin{equation}\label{eq18:othnoderr}
\begin{split}
        &e_m(n)  = d_m(n) - y_k(n)*s_{mk}(n)- \\ 
        & \sum_{l=1,l\neq k}^Ky_l(n)*s_{ml}(n), \; (m=1,\cdots,K,\; m \neq k).
\end{split}
\end{equation}
The local instantaneous cost function for other nodes can also be defined as \eqref{eq16:localcostfunc}. Since it is also correlated with $\mathbf{w}_k(n)$, to suppress the crosstalk effect, the partial differential for $\mathbf{w}_k(n)$ is considered, resulting in
\begin{equation}\label{eq19:partialdiff}
    \begin{split}
        \boldsymbol{\nabla}_{km} (n) &=\frac{\partial[e_m(n)]^2}{\partial\mathbf{w}_k(n)} \\&= -2e_m(n)[\mathbf{x}(n)*\hat{s}_{mk}(n)], (m \neq k).  
    \end{split}
\end{equation}
Moreover, substituting \eqref{eq10:compensate} into \eqref{eq19:partialdiff} yields 
\begin{equation}\label{eq20:othernodegradient}
    \boldsymbol{\nabla}_{km} (n) = -2 e_m(n)[\mathbf{x}(n)*\hat{s}_{mm}(n)]*c_{mk}(n), (m \neq k),
\end{equation}
where $e_m(n)$ and $\hat{s}_{mm}(n)$ denote the error signals and self-secondary path of the $m$th node, respectively. According to \eqref{eq17:localgradient},  \eqref{eq20:othernodegradient} can be simplified as
\begin{equation}\label{eq21:othernodegradient}
    \boldsymbol{\nabla}_{km} (n) = \boldsymbol{\nabla}_m(n)*c_{mk}(n), (m \neq k),
\end{equation}
where $\boldsymbol{\nabla}_m(n)$ represents the $m$th node's local gradients. The $k$th node can obtain these local gradients through a communication network. The node's gradient is mainly accountable for reducing the noise within the node, while the gradients received from other nodes are essential for suppressing interference. Therefore, the global control filter of the $k$th node can be updated as 
\begin{equation}\label{eq22:GCF}
    \begin{split}
            \mathbf{w}_k(n &+1) = \mathbf{w}_k(n) \\& - \mu\big[ \underbrace{\boldsymbol{\nabla}_k(n)}_{\text{Local~gradient}} + \underbrace{\sum_{m=1,m\neq k}^K \boldsymbol{\nabla}_m(n)*c_{mk}(n)}_{\text{Other~gradients}}\big],   
    \end{split}
\end{equation}
in which $\mu$ stands for the step size.

\begin{figure}[!t]
    \centering
    \includegraphics[width = 0.9\columnwidth]{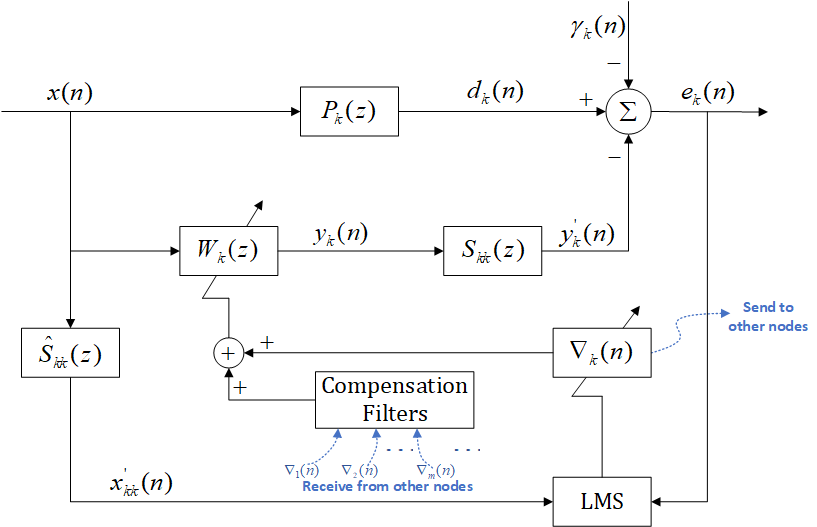}
    \caption{The block diagram of mixed-gradient distributed FxLMS (MGDFxLMS) algorithm for the $k$th node, where $P_k(z)$ represents the primary path and $\gamma_k(n)$ denotes the interference generated by other nodes.}
    \label{fig5:Distributedfornode}
\end{figure}

Figure~\ref{fig5:Distributedfornode} illustrates the block diagram of this proposed mixed-gradient distributed FxLMS (MGDFxLMS) algorithm. In the algorithm, each node computes the local gradient by using its own error signal and self-secondary path estimate, as shown in \eqref{eq17:localgradient}.  Subsequently, the local gradient is transmitted to other nodes. Once the node receives other nodes' gradients, it will utilize them to update the global control filter, as shown in \eqref{eq22:GCF}.  It is worth noting that by suppressing the interactive interference of secondary sources,  the proposed algorithm would achieve better noise reduction performance than standard decentralized methods. In addition, using the compensations filter eliminates the requirement for symmetric deployment of the system, unlike previous DMCANC methods. 


\subsection{Practical strategy overcoming the communication delay}

From the above descriptions, we can figure out that communication between different nodes plays a critical role in the noise reduction performance of the proposed method. However, no matter what type of communication method, wire or wireless communication, the communication delay issue in the real-world scenario cannot be circumvented and undoubtedly affects the performance and stability of the distributed system.

To analyze the impact of communication delay on the proposed method, we introduce a transmission delay sample, denoted as $\Delta$, into \eqref{eq22:GCF}:
\begin{equation}\label{eq23:delayGCF}
    \begin{split}
           \mathbf{w}_k(n&+1) = \mathbf{w}_k(n)\\& - \mu\big[\boldsymbol{\nabla}_k(n) + \sum_{m=1,m\neq k}^K\boldsymbol{\nabla}_m(n-\Delta)*c_{mk}(n)\big]. 
    \end{split} 
\end{equation}
It indicates that the $k$th node receives the local gradient sent by the other nodes from the previous $\Delta$ samples. Communication delays caused by network fluctuations prevent DMCANC from receiving information promptly, which may make the system unstable or even divergent, affecting the overall noise reduction effect. More detailed analysis will be explained in \ref{sec:analysis}. 

Like other adaptive algorithms, the delay in the system also decreases the maximum step size of the DMCANC algorithm. Hence, to avoid the effects of communication delays, a practical strategy that automatically shrinks the step size value is applied as
\begin{equation}\label{eq24:vss}
    \mu(n) = \mu_0e^{-2\Delta/f},
\end{equation}
where $\mu_0$ and $f$ represent the initial step size and sampling frequency, respectively. The delayed sample $\Delta$ can be obtained indirectly with the difference between the timestamps of sending and receiving. Hence, the update equation for each node becomes
\begin{equation}\label{eq25:VSSlocalgradient}
    \begin{split}
           &\mathbf{w}_k(n+1) = \mathbf{w}_k(n)\\&- \mu(n)\big[\boldsymbol{\nabla}_k(n) + \sum_{m=1,m\neq k}^K\boldsymbol{\nabla}_m(n-\Delta)*c_{mk}(n)\big].
    \end{split}
\end{equation}

From \eqref{eq24:vss}, it can be seen that the step size is reduced to ensure the stability of the system. In contrast, if the communication condition is fine, the step size is close to the initial step size and achieves faster convergence. For the case where each node has a different transmission delay, to ensure the stability of the system, $\Delta$ in \eqref{eq24:vss} takes the most significant delay among the received nodes, and hence,
\begin{equation}\label{eq:maxdelay}
    \Delta = \max[\{\Delta_m \mid m=1,2,\cdots,K \;\text{and} \; m\neq k \}],
\end{equation}
in which $\Delta_m$ stands for the delayed samples of the $m$th node. Therefore, a more general adaptive algorithm is thus derived as:
\begin{equation}\label{eq:Genearlupdate}
    \begin{split}
           &\mathbf{w}_k(n+1) = \mathbf{w}_k(n)\\&- \mu(n)\big[\boldsymbol{\nabla}_k(n) + \sum_{m=1,m\neq k}^K\boldsymbol{\nabla}_m(n-\Delta_m)*c_{mk}(n)\big]. 
    \end{split}
\end{equation}
\begin{table}[!t]
    \centering
    \caption{Pseudo-code of the ASSS-MGDFxLMS algorithm}\label{tab:algorithm}
    \begin{tabular}{l}
    \hline
    \textbf{Algorithm:} the ASSS-MGDFxLMS algorithm for the $k$th ANC nodes.\\
    \hline
    \textbf{Initialization:} Obtain estimated self-secondary path, $\hat{s}_{kk}(n)$, \\and compensation filters, $\mathbf{c}_{mk}(n)$.\\
    \textbf{Input:} The reference signal $x(n)$; The error signal $e_k(n)$; \\Received gradients from other nodes $\boldsymbol{\nabla}_m(n-\Delta_m)$, ($m\neq k$).\\
    \textbf{Output:} The control signal $y_k(n)$; The $k$th local gradients $\boldsymbol{\nabla}_k(n)$.\\
    \textbf{While} True \textbf{do}\\
     /*Combine gradients to obtain global control filter */\\
     ~~~$\Delta\leftarrow\max[\{\Delta_m \mid m\neq k \}]$\\
     ~~~$\mu(n) \leftarrow \mu_0e^{-2\Delta/f}$\\
     ~~~$\mathbf{w}_k(n+1) \leftarrow \mathbf{w}_k(n) + \mu(n)\cdot$ \\
     ~~~~~~~~~~~~~~~~~~~$\big[\boldsymbol{\nabla}_k(n) + \sum_{m=1,m\neq k}^K\boldsymbol{\nabla}_m(n-\Delta_m)*c_{mk}(n)\big]$\\
     /*Output control signal for the secondary source*/\\
     ~~~$y_k(n) \leftarrow \mathbf{w}_k^\mathrm{T}(n)\mathbf{x}(n) $\\
     /*Calculate local gradient and send to other nodes*/\\
     ~~~$\boldsymbol{\nabla}_k(n) \leftarrow [\mathbf{x}(n)*\hat{s}_{kk}(n)]\cdot e_k(n) $  ~~~~~~~~ $\triangleright$ Send to other nodes\\
     \textbf{end while}\\
    \hline
    \end{tabular}
\end{table}
The pseudo-code of the proposed auto-shrink step size MGDFxLMS (ASSS-MGDFxLMS) algorithm is elaborated in Table.~\ref{tab:algorithm}.
In the algorithm, the local gradient of each node is transmitted and combined with the compensation filter to compute the global control filters. Furthermore, the auto-shrink step size strategy allows the step size value to be adaptive to communication delays to enhance the system's robustness. Compared to other methods of exchanging local control filters, dealing with such a situation is more flexible by transmitting local gradients.

\subsection{Computational complexity}
The DMCANC framework distributes the substantial computational load of the centralized MCANC algorithm across multiple processors. In Table.~\ref{tab2:computation}, we compare the computations required by various algorithms on a single processor. As expected, MCFxLMS demands more computation because a single processor must handle all inputs and outputs. In contrast, distributed algorithms, supported by multiple processors, share the computational burden, thereby reducing the load per processor. When examining different distributed algorithms, we find that the proposed MGDFxLMS algorithm offers a moderate computational effort per processor, and its ASSS strategy does not increase too many computational requirements. Additionally, because MGDFxLMS employs convolution operations to obtain global control filters, it can benefit from frequency-domain implementations to further decrease its computational cost. Overall, by integrating multiple low-cost processors, the proposed algorithm achieves performance comparable to the centralized approach at a lower per-processor computational effort.

\begin{table}[!t]
    \centering
    \caption{Computational complexity of different algorithms on a single processor}
    \begin{tabular}{ccc}
    \hline
         & Multiplication & Addition \\
         \hline
       \multirow{2}{7em}{MCFxLMS\cite{Elliott1987MEANC}} & $K^2(2N+L)$ & $K^2(N+L-1)$ \\
                                  &{$+KN$}  & {$+ K(N-1)$}\\     
        \multirow{2}{7em}{DFxLMS\cite{Chu2019DiffusionANC}} & \multirow{2}{6em}{$(K+3)N+L$} & $(K+1)N$\\
                                  &     & $+L-2$ \\
        \multirow{2}{7em}{ADFxLMS\cite{Li2023Distributed}} & $(K+1)^2N$ & $(K^2+1)N$ \\
                                   &  $+KL$  &  $+K(L-1)-1$ \\
        \multirow{2}{7em}{MGDFxLMS} & $L + (3+H)N$ & $(K+1)N + L$\\
                                    &   $- H(H-3)-2$   & $+ (H-1)(N-H+1) -2$ \\
        \multirow{2}{7em}{ASSS-MGDFxLMS} & $L + (3+H)N$  &  $(K+1)N + L $ \\
                            & $- H(H-3)$  &  $+ (H-1)(N-H+1) -2$ \\
        \hline
        \hline
         \multicolumn{3}{c}{$K = 6, N = 512, L = 256, H = 33$} \\
         \hline
         MCFxLMS\cite{Elliott1987MEANC} & $49152$ & $33750$ \\
        DFxLMS\cite{Chu2019DiffusionANC} & $4864$ & $3838$ \\
        ADFxLMS\cite{Li2023Distributed} & $26624$ & $20473$ \\
        MGDFxLMS & {$17696$} & {$19198$} \\
        ASSS-MGDFxLMS & {$17698$}  &  {$19198$}  \\
        \hline
        \multicolumn{3}{l}{{*$K$ is the number of nodes.}}\\
        \multicolumn{3}{l}{{*$N,L,H$ represent the length of control filters, estimated secondary paths}} \\
        \multicolumn{3}{l}{{and compensation filters, respectively.}} \\
    \end{tabular}
    \label{tab2:computation}
\end{table}

\section{Performance Analysis}\label{sec:analysis}
This section examines the steady-state performance of the MGDFxLMS algorithm and its convergence behavior in two scenarios: the ideal circumstance and the communication latency issue.

\subsection{Optimal global control filter}
To achieve global noise reduction, the DMCANC should have the same cost function as the conventional one as expressed in \eqref{eq3:costfunc}. By substituting \eqref{eq2:error} and \eqref{eq1:controlsignal} into \eqref{eq3:costfunc}, we can obtain:
\begin{equation}\label{eq26:costfunc1}
    J = \sum_{m=1}^{K}\mathbb{E}[d_{m}(n) - \sum_{k=1}^{K}\mathbf{w}_k^\mathrm{T}(n)\mathbf{x}(n)*{s}_{mk}(n)]^2.
\end{equation}
Under the assumption of slow updating, \eqref{eq26:costfunc1} can be further rewritten as
\begin{equation}\label{eq27:costfunc2}
    J = \sum_{m=1}^{K}\mathbb{E}[(d_{m}(n) - \sum_{k=1}^{K}\mathbf{x'}_{km}^\mathrm{T}(n)\mathbf{w}_k(n))^2].
\end{equation}
Expanding \eqref{eq27:costfunc2} yields
\begin{equation}\label{eq28:costfunc3}
    J = \sum_{m=1}^{K}[\sigma_{d_m}^2-2\sum_{k=1}^K\mathbf{P}_{km}^\mathrm{T}\mathbf{w}_k(n)+\sum_{k=1}^K\sum_{l=1}^K\mathbf{w}^\mathrm{T}_k(n)\mathbf{R}_{kl,m}\mathbf{w}_l(n)],
\end{equation}
where
\begin{equation}\label{eq29:definition}
    \begin{cases}
        \sigma_{d_m}^2 & = \; \mathbb{E}[d^2_m(n)],\\
        \mathbf{P}_{km} & = \; \mathbb{E}[d_m(n)\mathbf{x'}_{km}(n)],\\
        \mathbf{R}_{kl,m} & = \; \mathbb{E}[\mathbf{x'}_{km}(n)\mathbf{x'}^\mathrm{T}_{lm}(n)].
    \end{cases}
\end{equation}
By taking the partial differential of \eqref{eq28:costfunc3} with respect to the $k$th global control filter $\mathbf{w}_k(n)$, we can get
\begin{equation}\label{eq30:partialdiffer}
    \frac{\partial J}{\partial \mathbf{w}_k(n)} = 2\sum_{m=1}^K[\sum_{l=1}^K\mathbf{R}_{kl,m}\mathbf{w}_l(n)-\mathbf{P}_{km}].
\end{equation}
To derive the optimal global control filter, \eqref{eq30:partialdiffer} is set to zero, resulting in
\begin{equation}\label{eq31:optGCF}
    \sum_{m=1}^K\mathbf{R}_{kk,m} \cdot \mathbf{w}_{k,opt} = \sum_{m=1}^K [\mathbf{P}_{km} - \sum_{l=1,l\neq k}^K\mathbf{R}_{kl,m}\mathbf{w}_l(n)].  
\end{equation}
Therefore, the optimal solution for the $k$th global control filter can be deduced as
\begin{equation}\label{eq32:optkGCF}
    \mathbf{w}_{k,opt} = \sum_{m=1}^K\{\mathbf{R}^{-1}_{kk,m} \cdot [\mathbf{P}_{km} - \sum_{l=1,l\neq k}^K\mathbf{R}_{kl,m}\mathbf{w}_l(n)]\}.  
\end{equation}
From \eqref{eq32:optkGCF}, it can be found that other nodes' global control filters affect the kth node's optimal solution, exhibiting in coupling phenomenon  \cite{Hansen1997ACNV}.

\subsection{Convergence analysis without communication delay}
In an ideal case, each node can receive other nodes' information in time at each iteration, resulting in no communication delay in the distributed network. Therefore, the global control filter for each node can be updated with the instant gradient from other nodes as expressed in \eqref{eq22:GCF}. According to \eqref{eq5:filteredx}, \eqref{eq10:compensate} and \eqref{eq17:localgradient}, it can be rewritten as
\begin{equation}\label{eq33:GCFupdate}
    \begin{split}
           \mathbf{w}_k(n+1) = \mathbf{w}_k(n) &+ \mu \mathbf{x'}_{kk}(n)e_k(n)\\ &+ \mu \sum_{m=1,m\neq k}^K \mathbf{x'}_{km}(n)e_m(n).  
    \end{split} 
\end{equation}

Moreover, we defined the weight error at the $k$th node as
\begin{equation}\label{eq34:GCFdifftoideal}
    \mathbf{v}_k(n) = \mathbf{w}_{k,opt} - \mathbf{w}_k(n). 
\end{equation}
Hence, \eqref{eq33:GCFupdate} can be rewritten as
\begin{equation}\label{eq35:GCFdiff}
    \mathbf{v}_k(n+1) = \mathbf{v}_k(n) - \mu \sum_{m=1}^K \mathbf{x'}_{km}(n)e_m(n).  
\end{equation}
Based on \eqref{eq34:GCFdifftoideal}, $e_m(n)$ can be represented as
\begin{equation}\label{eq36:errkm}
        e_m(n) = d_m(n) - \sum_{l=1}^K  \mathbf{x'}_{lm}^{\mathrm{T}}(n)[\mathbf{w}_{l,opt} - \mathbf{v}_l(n)].
\end{equation}
Substituting \eqref{eq36:errkm} into \eqref{eq35:GCFdiff} yields
\begin{equation}\label{eq37:GCFdiff1}
    \begin{split}
        &\mathbf{v}_k(n+1) = \mathbf{v}_k(n) - \mu \sum^K_{m=1}\mathbf{x'}_{km}(n)[d_m(n) - \\
       &\mathbf{x'}_{km}^\mathrm{T}(n)(\mathbf{w}_{k,opt}-\mathbf{v}_k(n)) - \sum_{l=1\neq k}^K \mathbf{x'}_{lm}^\mathrm{T}(\mathbf{w}_{l,opt}-\mathbf{v}_l(n))].
    \end{split}
\end{equation}
By taking the expectation of \eqref{eq37:GCFdiff1}, it becomes
\begin{equation}\label{eq38:GCFdiff2}
    \begin{split}
        \mathbf{v}_k(n+1) =  \mathbf{v}_k(n) &- \mu \sum^K_{m=1}[\mathbf{P}_{km} \\&- \mathbf{R}_{kk,m}(\mathbf{w}_{k,opt}-\mathbf{v}_k(n)) 
        \\&-\sum_{l=1,\neq k}^K \mathbf{R}_{kl,m}(\mathbf{w}_{l,opt}-\mathbf{v}_l(n))].
    \end{split}
\end{equation}

According to \eqref{eq34:GCFdifftoideal}, \eqref{eq31:optGCF} can be rewritten as 
\begin{equation}\label{eq39:optGCF1}
\begin{split}
    \sum_{m=1}^K\mathbf{R}_{kk,m} \cdot \mathbf{w}_{k,opt} &= \sum_{m=1}^K [\mathbf{P}_{km} \\&- \sum_{l=1,\neq k}^K\mathbf{R}_{kl,m}(\mathbf{w}_{l,opt}-\mathbf{v}_l(n))]. 
\end{split}
\end{equation}
Substituting \eqref{eq39:optGCF1} into \eqref{eq38:GCFdiff2} yields
\begin{equation}\label{eq40:GCFdiff3}
        \mathbf{v}_k(n+1) =  \mathbf{v}_k(n) - \mu \sum_{m=1}^K\mathbf{R}_{kk,m}\mathbf{v}_{k}(n).
\end{equation}

Since the auto-correlation matrix $\mathbf{R}_{kk,m}$ is symmetric, it can be decomposed through the orthogonal transformation as
\begin{equation}\label{eq41:orthtransform}
        \boldsymbol{\Lambda}_{k,m} = \mathbf{Q}^\mathrm{T}\mathbf{R}_{kk,m}\mathbf{Q},
\end{equation}
where $\mathbf{Q}$ represents the orthogonal matrix, and hence, 
\begin{equation}\label{eq42:orthmatrix}
       \mathbf{Q}\mathbf{Q}^\mathrm{T} = \mathbf{I}.
\end{equation}
In the above equation, $\mathbf{I}$ is the identity matrix, and $\boldsymbol{\Lambda}_{k,m}$ in \eqref{eq41:orthtransform} represents the diagonal eigenvalue matrix of $\mathbf{R}_{kk,m}$ given by
\begin{equation}\label{eq43:eigenvalue}
        \boldsymbol{\Lambda}_{k,m} = \text{diag}[\lambda_{km,1}, \lambda_{km,2}, \cdots,\lambda_{km,N}]
\end{equation}
where $\lambda_{km,i}$ denotes the $i$th eigenvalue.
By multiplying the transpose of orthogonal matrix on both side, \eqref{eq40:GCFdiff3} can be rewritten as 
\begin{equation}\label{eq44:transGCFdiffer}
    \begin{split}
            \mathbf{Q}^\mathrm{T}\mathbf{v}_k(n+1) = &\mathbf{Q}^\mathrm{T}\mathbf{v}_k(n)\\ &- \mu \sum_{m=1}^K\mathbf{Q}^\mathrm{T}\mathbf{R}_{kk,m}\mathbf{Q}\mathbf{Q}^\mathrm{T}\mathbf{v}_{k}(n).    
    \end{split}
\end{equation}
According to \eqref{eq41:orthtransform}, \eqref{eq44:transGCFdiffer} can be simplified as 
\begin{equation}\label{eq45:transGCFdiffer1}
    \mathbf{v}'_k(n+1) = \mathbf{v}'_k(n) - \mu \sum_{m=1}^K\boldsymbol{\Lambda}_{k,m}\mathbf{v}'_{k}(n),
\end{equation}
where
\begin{equation}\label{eq46:GCFdiffer1}
    \mathbf{v}'_k(n) = \mathbf{Q}^\mathrm{T}\mathbf{v}_k(n).
\end{equation}

Hence, the $i$th element of \eqref{eq45:transGCFdiffer1} can be found as  
\begin{equation}\label{eq47:transGCFdifferi}
    {v}'_{k,i}(n+1) = {v}'_{k,i}(n) (1 - \mu \sum_{m=1}^K\lambda_{km,i} ).
\end{equation}
To guarantee the convergence of the algorithm, we can derive to 
\begin{equation}\label{eq48:requirements}
    |1 - \mu \sum_{m=1}^K\lambda_{km,i}| < 1,
\end{equation}
which can be deduced as
\begin{equation}\label{eq49:mubound}
  0 < \mu < \frac{2}{\sum_{m=1}^K\lambda_{km,i}}.
\end{equation}
For the purpose of global stability, the maximum eigenvalue, $\lambda_{km,max}$, of $\mathbf{R}_{kk,m}$ should be taken into account. Hence, the step size bound to ensure the convergence of the algorithm under no communication delay should be
\begin{equation}\label{eq50:mubound1}
  0 < \mu < \frac{2}{\sum_{m=1}^K\lambda_{km,max}}, 
\end{equation}
which is the same as the conventional centralized MCFxLMS algorithm \cite{Elliott2001SPAC}.

\subsection{Convergence analysis under communication delays}
In practice, the unstable network results in delayed information transmission. 
For simplicity, each node is assumed to own the same delay samples, and, hence,  it receives a delayed gradient from previous $\Delta$ samples as
\begin{equation}\label{eq51:delayedGCF}
    \begin{split}
        \mathbf{w}_k(n+1) = &\mathbf{w}_k(n) + \boldsymbol{\nabla}_k(n-\Delta)\\ &+ \sum_{m=1,m\neq k}^K\boldsymbol{\nabla}_m(n-\Delta)*c_{mk}(n).  
    \end{split}
\end{equation}
Following the same derivation as the previous section, we can obtain
\begin{equation}\label{eq52:delayedtransGCFdiffer1}
    \mathbf{v}'_k(n+1) = \mathbf{v}'_k(n) - \mu \sum_{m=1}^K\boldsymbol{\Lambda}_{k,m}\mathbf{v}'_{k}(n-\Delta).
\end{equation}
Hence, the $i$th element of \eqref{eq52:delayedtransGCFdiffer1} is given by
\begin{equation}\label{eq53:delayedtransGCFdifferi}
    {v}'_{k,i}(n+1) = {v}'_{k,i}(n) - \mu \sum_{m=1}^K{\lambda}_{km,i}{v}'_{k,i}(n-\Delta),
\end{equation}
and their the single-sided $z$-transform is derived as 
\begin{equation}\label{eq54:zdomain}
\begin{split}
    &z{V}'_{k,i}(z) - z{v}'_{k,i}(0)  =  {V}'_{k,i}(z)\\ &-\mu \sum_{m=1}^K{\lambda}_{km,i} [z^{-\Delta}{V}'_{k,i}(z) +z^{-\Delta}\sum_{p=1}^{\Delta}v_{k,i}(-p)z^p].
\end{split}
\end{equation}
Since the system is causal, $v'_{k,i}(n) = 0$ for all $n<0$, and, hence, \eqref{eq54:zdomain} is derived as 
\begin{equation}\label{eq55:zdomaineq}
    V'_{k,i}(z) = \frac{z^{\Delta+1}v'_{k,i}(0)}{z^{\Delta+1}-z^{\Delta}+\mu\sum_{m=1}^K{\lambda}_{km,i}},
\end{equation}
whose characteristic equation is found as
\begin{equation}\label{eq56:characteq}
   {z^{\Delta+1}-z^{\Delta}+\mu\sum_{m=1}^K{\lambda}_{km,i}} = 0.
\end{equation}

Based on the property of the root locus, it is evident that when the parameter $\mu$ increases, the magnitude of $z$ will also increase proportionally. To ensure system stability, the poles of equation \eqref{eq55:zdomaineq} or the roots of equation \eqref{eq56:characteq} must lie within the unit circle, meaning that their absolute value must be less than or equal to 1, i.e., $|z|\leq 1$. As a result, equating the absolute value of $z$ to 1 will yield the maximum value of $\mu$.

According to Euler's formula, we can get 
\begin{equation}\label{eq57:zejalpha}
   z=e^{j\alpha} = \cos{\alpha} + j\sin{\alpha}.
\end{equation}
Given that $\mathbf{R}_{kk,m}$ is a real symmetric matrix, it follows that its eigenvalue $\lambda_{km,i}$ must also be a real number. Therefore, by replacing \eqref{eq57:zejalpha} with \eqref{eq56:characteq}, we obtain
\begin{equation}\label{eq58:zequation}
   \begin{cases}
       \cos{(\Delta+1)\alpha} - \cos{\Delta \alpha} + \mu\sum_{m=1}^K{\lambda}_{km,i} = 0, \\
       \sin{(\Delta+1)\alpha} - \sin{\Delta \alpha} = 0
   \end{cases}
\end{equation}
The solution of \eqref{eq58:zequation} can be deduced as
\begin{equation}\label{eq59:solution}
\begin{cases}
    \alpha = \frac{\pi}{2\Delta +1}, \\
    \mu = \frac{2}{\sum_{m=1}^K{\lambda}_{km,i}} \sin{\frac{\pi}{2(2\Delta +1)}}.
\end{cases}
\end{equation}
Therefore, the step size bound of the proposed algorithm under communication delay should be
\begin{equation}\label{eq60:ssbounddelay}
    0<\mu < \frac{2}{\sum_{m=1}^K{\lambda}_{km,max}} \sin{\frac{\pi}{2(2\Delta +1)}}.
\end{equation}

By comparing \eqref{eq50:mubound1} and \eqref{eq60:ssbounddelay}, it can be found that communication delays reduce the step size boundary, making the convergence conditions more stringent. In this scenario, the proposed auto-shrink step size strategy will enhance the stability of the system.
   


\section{Numerical Simulations}\label{sec:sim}
\begin{figure}[!t]
    \centering
    \includegraphics[width = 0.9\columnwidth, height = 19.5cm]{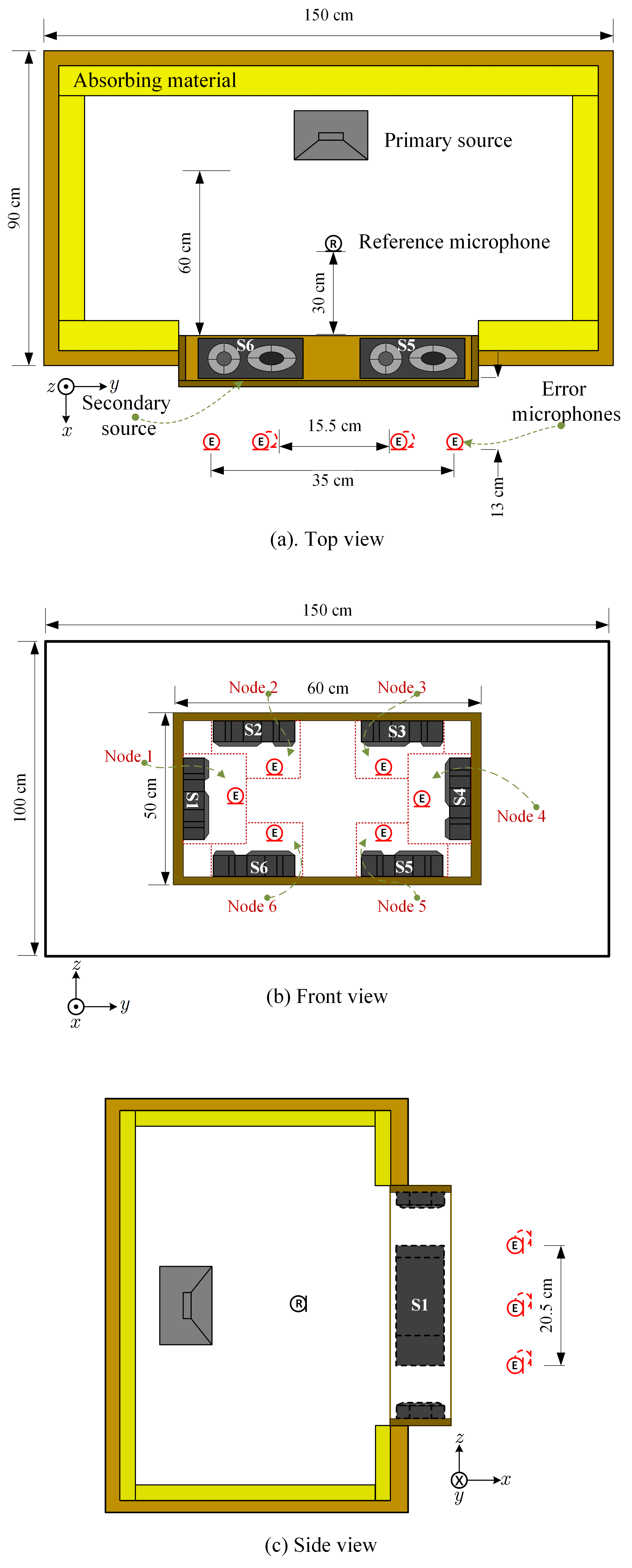}
    \caption{The schematic of a 6-node MCANC system installed in an open aperture of a noise chamber: (a) the top view of the chamber; (b) the front view of the chamber and the distribution of 6 nodes; (c) the side view of the chamber.}
    \label{fig:configuration}
\end{figure}

In this section, we first validated the performance of our proposed MGDFxLMS algorithm on an MCANC system with $6$ nodes. Subsequently, we investigated the noise reduction (NR) performance under communication delay to demonstrate the robustness of the proposed ASSS-MGDFxLMS algorithm. The primary and secondary paths are measured from a noise chamber with an ANC window \cite{shi2023computation}, as shown in Fig.~\ref{fig:configuration}. The length of the secondary paths, the compensation filters, and the global control filters are set to $256$, $33$, and $512$, respectively. The sampling frequency is chosen as $16,000$Hz. In these numerical simulations, we conducted a comparison between our proposed MGDFxLMS algorithm and the conventional centralized MCFxLMS algorithm \cite{Elliott1987MEANC}, DFxLMS \cite{Chu2019DiffusionANC}, and ADFxLMS \cite{Li2023Distributed}.

To evaluate the noise reduction (NR) performance of the algorithm, the normalized squared error (NSE) is applied, defined as
\begin{equation}\label{eq61:NSE}
    \text{NSE}(n) = 10\log_{10}(\frac{\mathbb{E}(e^2_m(n))}{\mathbb{E}(d^2_m(n))}),
\end{equation}
where the expectation value is calculated by averaging the signal over $5,000$ samples.

\subsection{Broadband noise cancellation} \label{subsec:case1}

\begin{figure}[!t]
    \centering
    \includegraphics[width = 0.9\columnwidth]{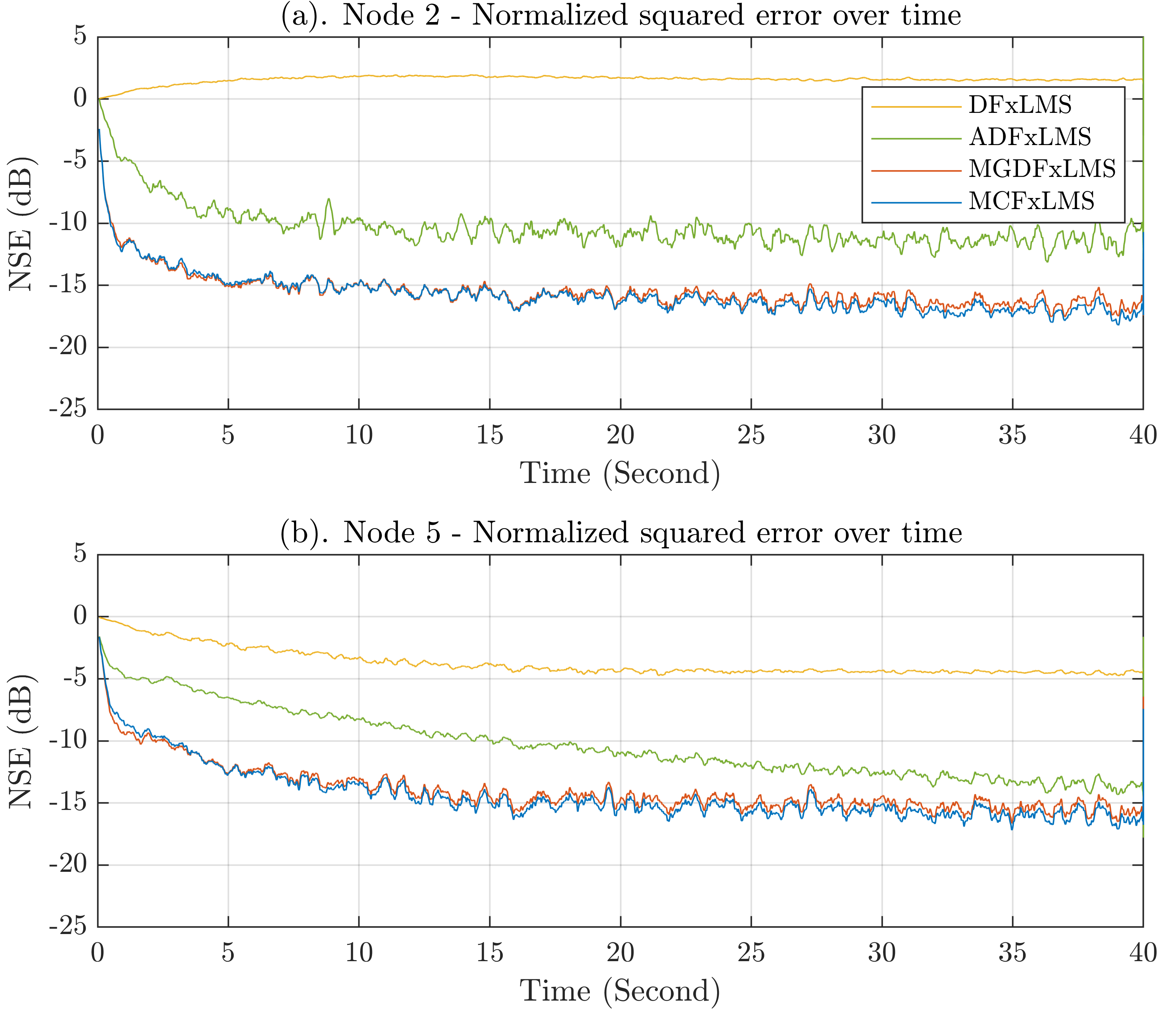}
    \caption{NR performances with different algorithms: (a) NSE over time of node 2; (b) NSE over time of node 5.}
    \label{fig:case1_1}
\end{figure}
\begin{figure}[!t]
    \centering
    \includegraphics[width = 0.9\columnwidth]{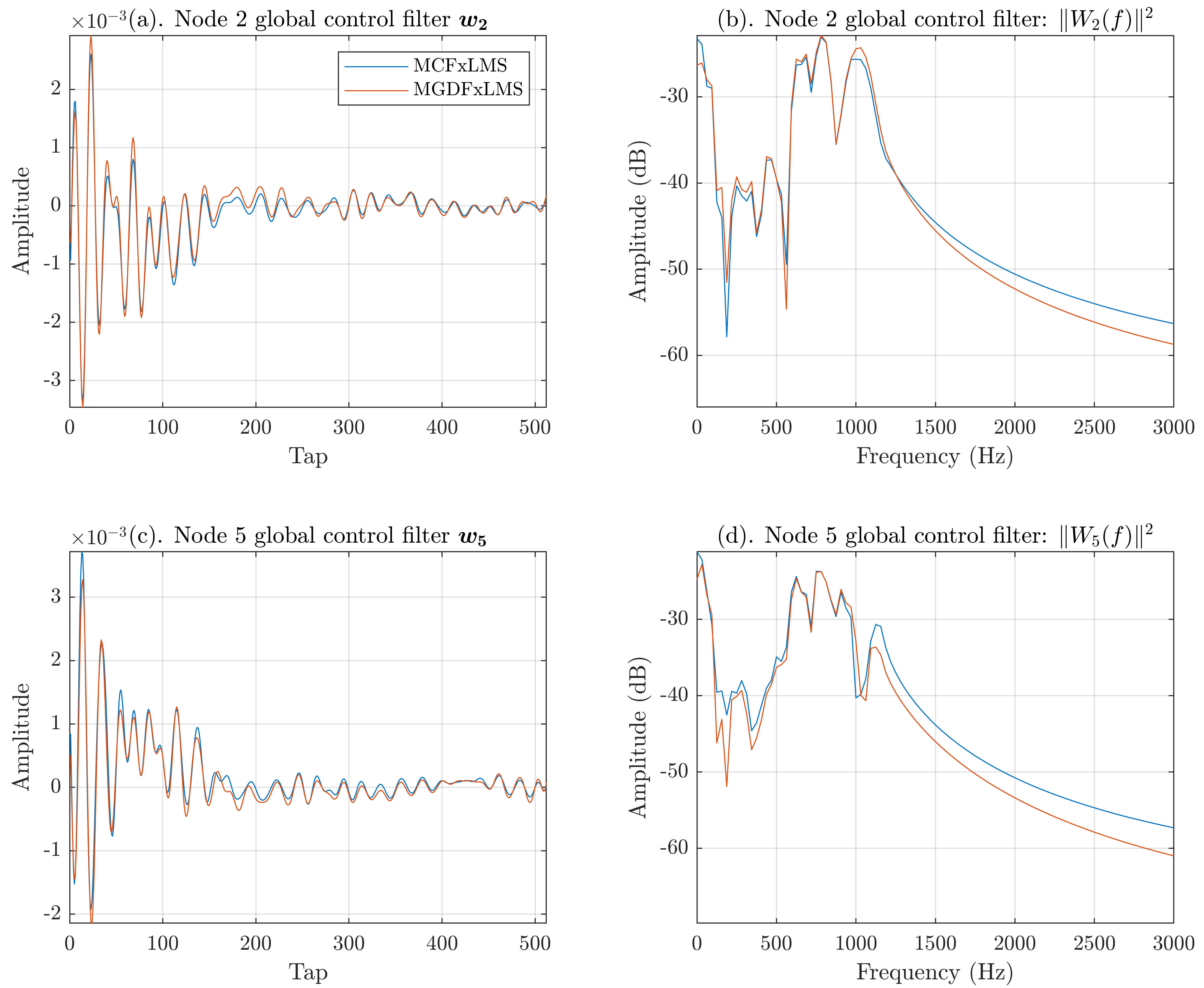}
    \caption{Comparison of the global control filters at steady state in the centralized MCFxLMS and proposed MGDFxLMS algorithm: (a) The global control filter weights of node 2; (b) Frequency response of global control filter in node 2; (c) The global control filter weights of node 5; (d) Frequency response of global control filter in node 5;}
    \label{fig:case1_2}
\end{figure}

To verify the algorithm, the primary noise taken is broadband noise with a frequency range of 100 to 1,000Hz. The MCFxLMS and proposed MGDFxLMS algorithms use a step size of $1\times10^{-6}$, whereas the ADFxLMS algorithm has a step size of $5\times10^{-6}$, and the DFxLMS algorithm uses a step size of $1\times10^{-7}$.
From Fig.~\ref{fig:case1_1}, the DFxLMS algorithm fails to converge. DFxLMS uses a simplistic topology-based combination rule to combine local control filters into global control filters. This strategy does not account for the differences in sound field information between nodes, which reduces the system's stability and renders it less effective in more complex scenarios.
The ADFxLMS algorithm incorporates the cross-secondary path in the update procedure, leading to better NR performance than the DFxLMS algorithm. However, its noise reduction performance at each node is unbalanced thus affecting the global noise reduction.
In contrast, the proposed MGDFxLMS algorithm has an almost similar expression to the traditional centralized MCFxLMS algorithm and thus presents the same noise reduction performance. Fig.~\ref{fig:case1_2} further illustrates that the proposed algorithm achieves identical global control filters with the centralized algorithm at the steady state. Therefore, the proposed algorithm shows satisfactory performance in broadband noise cancellation.

\subsection{{The effect of compensation filter lengths}}

\begin{figure}[!t]
    \centering
    \includegraphics[width = 0.9\columnwidth,height = 6.8cm]{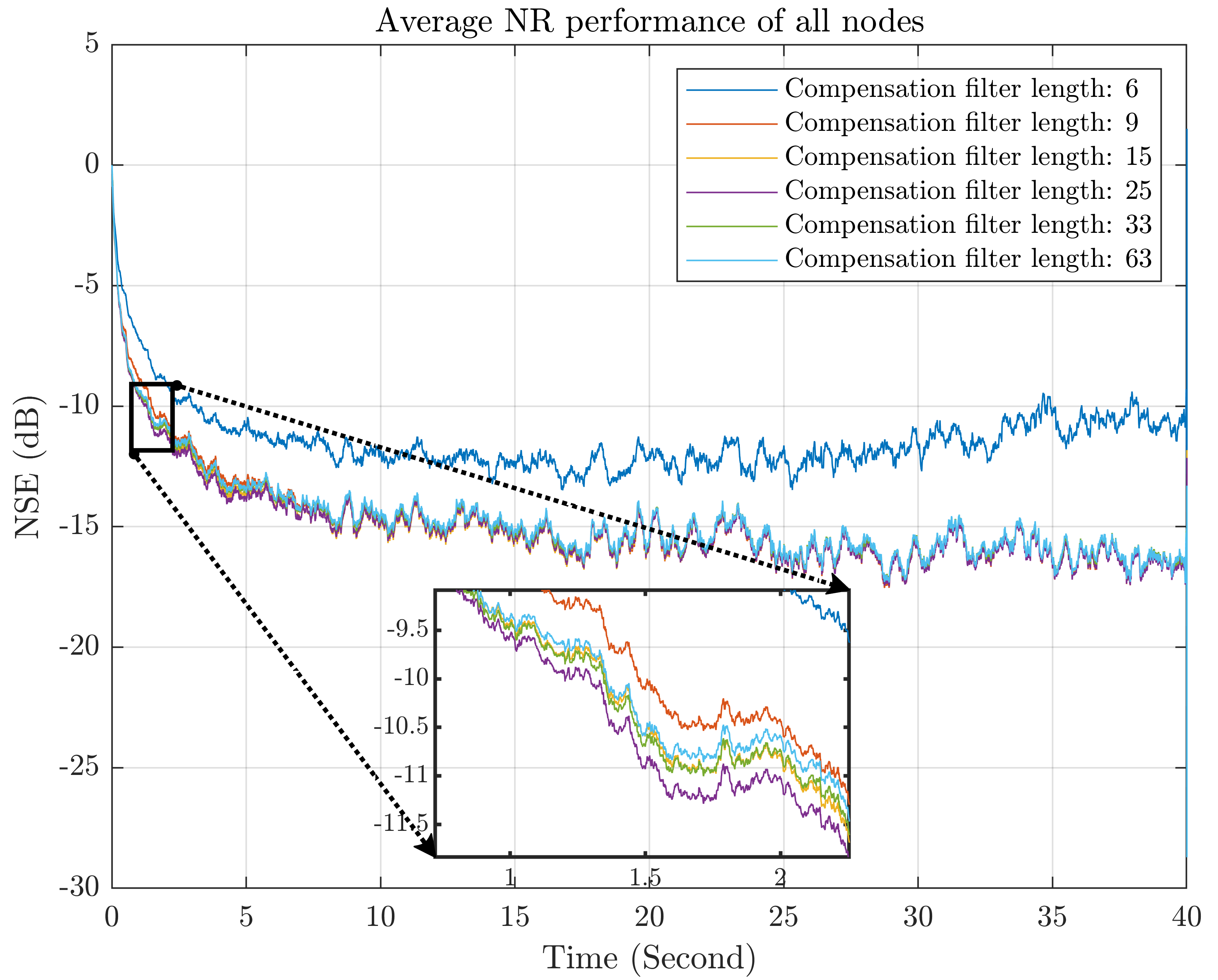}
    \caption{{Average NR performance of all nodes with different compensation filter lengths}}
    \label{fig:CFNR}
\end{figure}

{The proposed MGDFxLMS algorithm uses compensation filters to integrate gradient information from multiple nodes. To assess its effectiveness, we examined how varying the length of these filters influences the DMCANC system’s performance. As shown in Fig.~\ref{fig:CFNR}, filters that are too short cannot properly reconcile the differences between self and cross-secondary paths, causing the algorithm to diverge. By increasing the filter length, the system becomes stable, yielding similar noise reduction levels at steady state. However, the NSE slopes indicate that while adequately long filters ensure convergence, both overly short and excessively long filters negatively impact convergence speed. Notably, shorter filters lower computational costs, highlighting a trade-off between ANC performance and computational efficiency.}

\subsection{Real recorded noise cancellation}

\begin{figure}[!t]
    \centering
    \includegraphics[width = 0.9\columnwidth,height = 6.8cm]{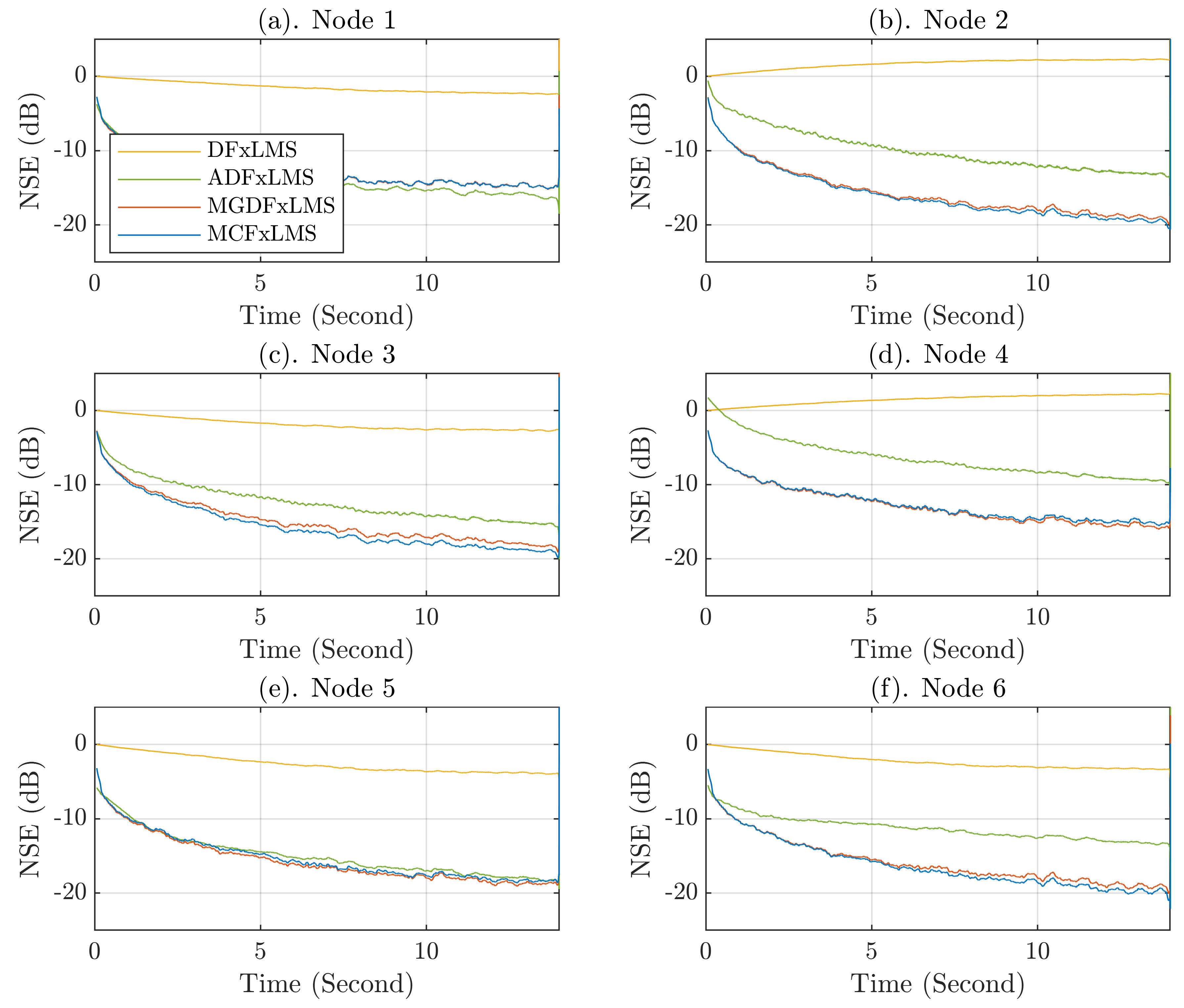}
    \caption{Recorded compressor NR performance with different algorithms: (a) - (f) NSE over time from node 1 to node 6.}
    \label{fig:case2_1}
\end{figure}

In this simulation, we chose a real-record compressor noise as the primary noise. For the centralized MCFxLMS, the proposed MGDFxLMS, ADFxLMS, and DFxLMS, the step size is selected as $5\times10^{-6}$, $5\times10^{-6}$, $3\times10^{-4}$, and $1\times10^{-7}$, respectively. 
Fig.~\ref{fig:case2_1} shows that the noise reduction performance of DFxLMS is worse and some nodes even diverge. The ADFxLMS technique presents different noise reduction performances for each node, highlighting the inadequacy of combined processing methods that solely rely on a weighted summing operation. 
However, the MGDFxLMS method, derived directly from the standard centralized algorithm, exhibits comparable noise reduction capabilities at each node and achieves higher global noise control performance. Hence, the proposed MGDFxLMS algorithm is applicable in practical scenarios including real noise sources and acoustic paths.

\begin{figure}[!t]
    \centering
    \includegraphics[width = 0.9\columnwidth,height = 6.5cm]{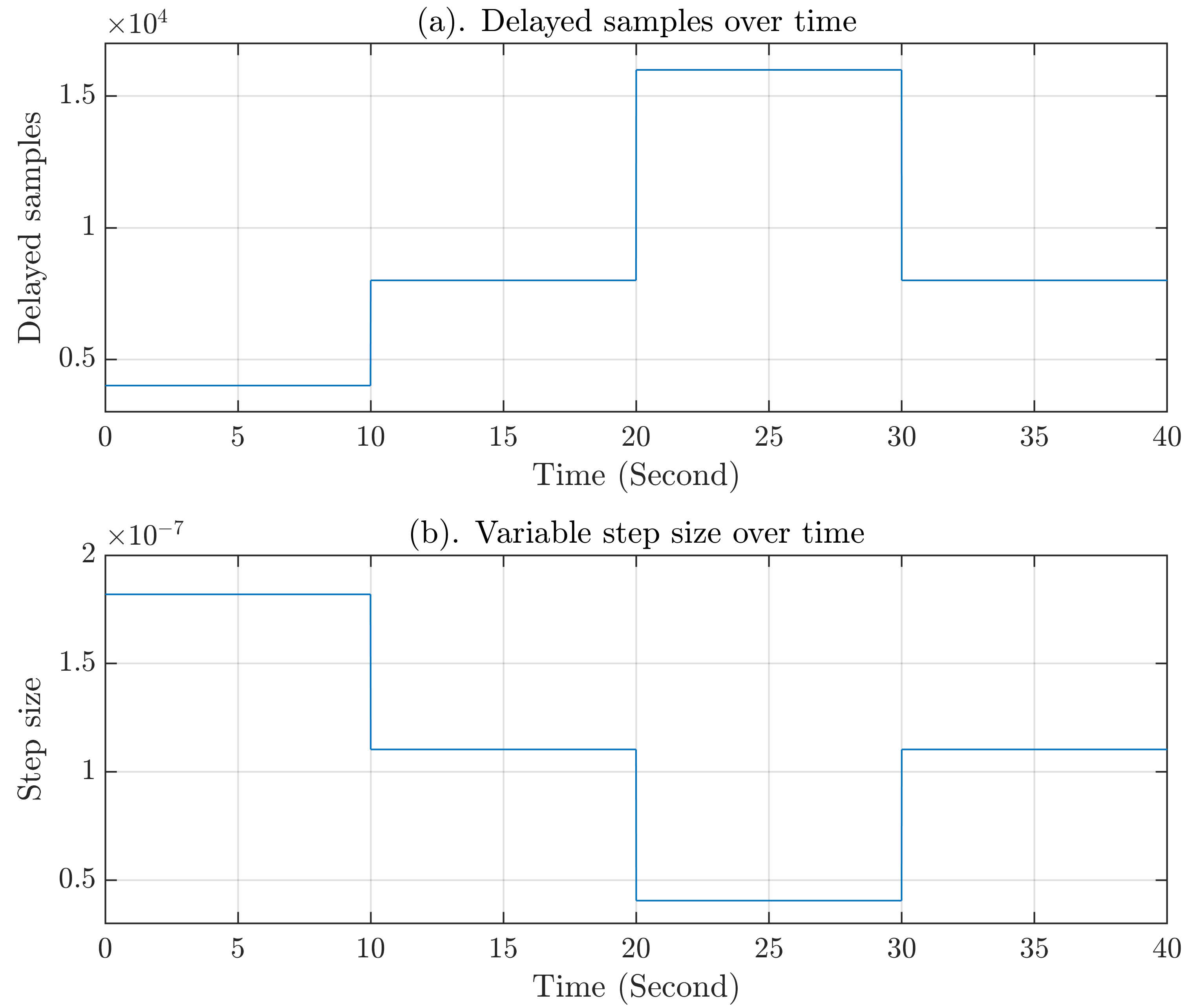}
    \caption{There exists communication delay in the distributed network: (a) Delayed samples over time. (b) Corresponding step size to the delayed samples in ASSS-MGDFxLMS algorithm.}
    \label{fig:case3_2}
\end{figure}

\subsection{Noise reduction performance of the algorithms under sudden changes in communication delay}

In this simulation, the communication delay between nodes is considered to verify the robustness of the proposed method.  The primary noise is the same as that used in Sec.~\ref{subsec:case1}, and the initial step size is chosen as $3\times10^{-7}$ for the ASSS-MGDFxLMS algorithm, and the rest is set as $1.5\times10^{-7}$.

\begin{figure}[!t]
    \centering
    \includegraphics[width = 0.9\columnwidth]{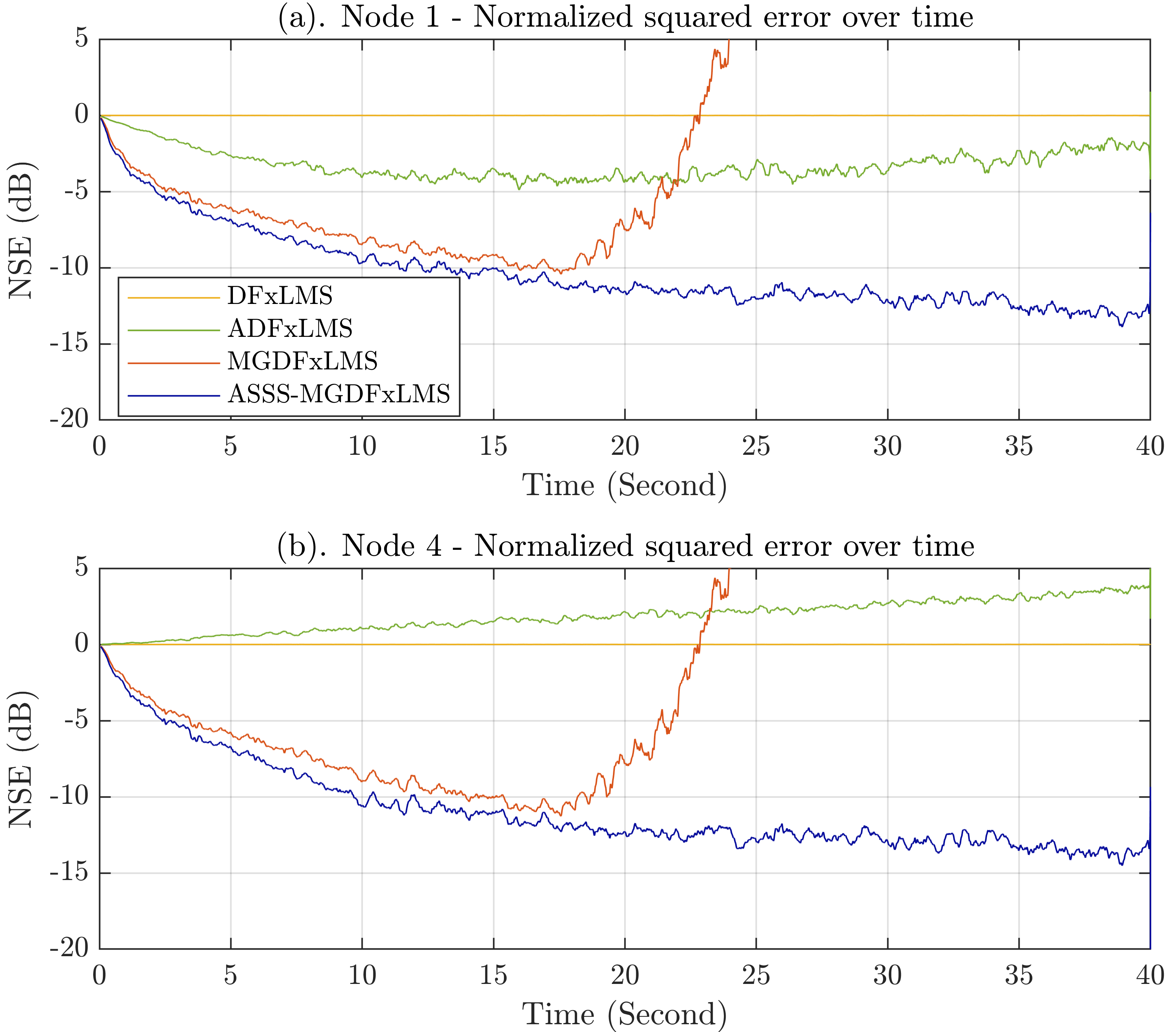}
    \caption{NR performance with different DMCANC algorithms when the communication delay suddenly changes: (a) NSE over time of node 1; (b) NSE over time of node 4;}
    \label{fig:case3_1}
\end{figure}

The communication delay undergoes variations at the 10s (from the initial 4,000 delayed samples to 8,000 samples), the 20s (from 8,000 samples to 16,000 samples), and the 30s (from 16,000 samples to 8,000 samples), as illustrated in Fig.~\ref{fig:case3_2} (a). The simulation results in Fig.~\ref{fig:case3_1} demonstrated that the DFxLMS, ADFxLMS, and MGDFxLMS without auto-shrink step size do not result in any noise reduction or even diverge as a result of the failure to receive information from other nodes in time. Conversely, the proposed ASSS-MGDFxLMS algorithm can guarantee algorithm convergence and simultaneously achieve a satisfactory noise reduction effect by utilizing the auto-shrink step size strategy to dynamically adjust the step size in accordance with the communication delay, as illustrated in Fig.~\ref{fig:case3_2}.

\subsection{Noise reduction performance of the proposed ANC system under fluctuating network}

\begin{figure}[!t]
    \centering
    \includegraphics[width = 0.9\columnwidth]{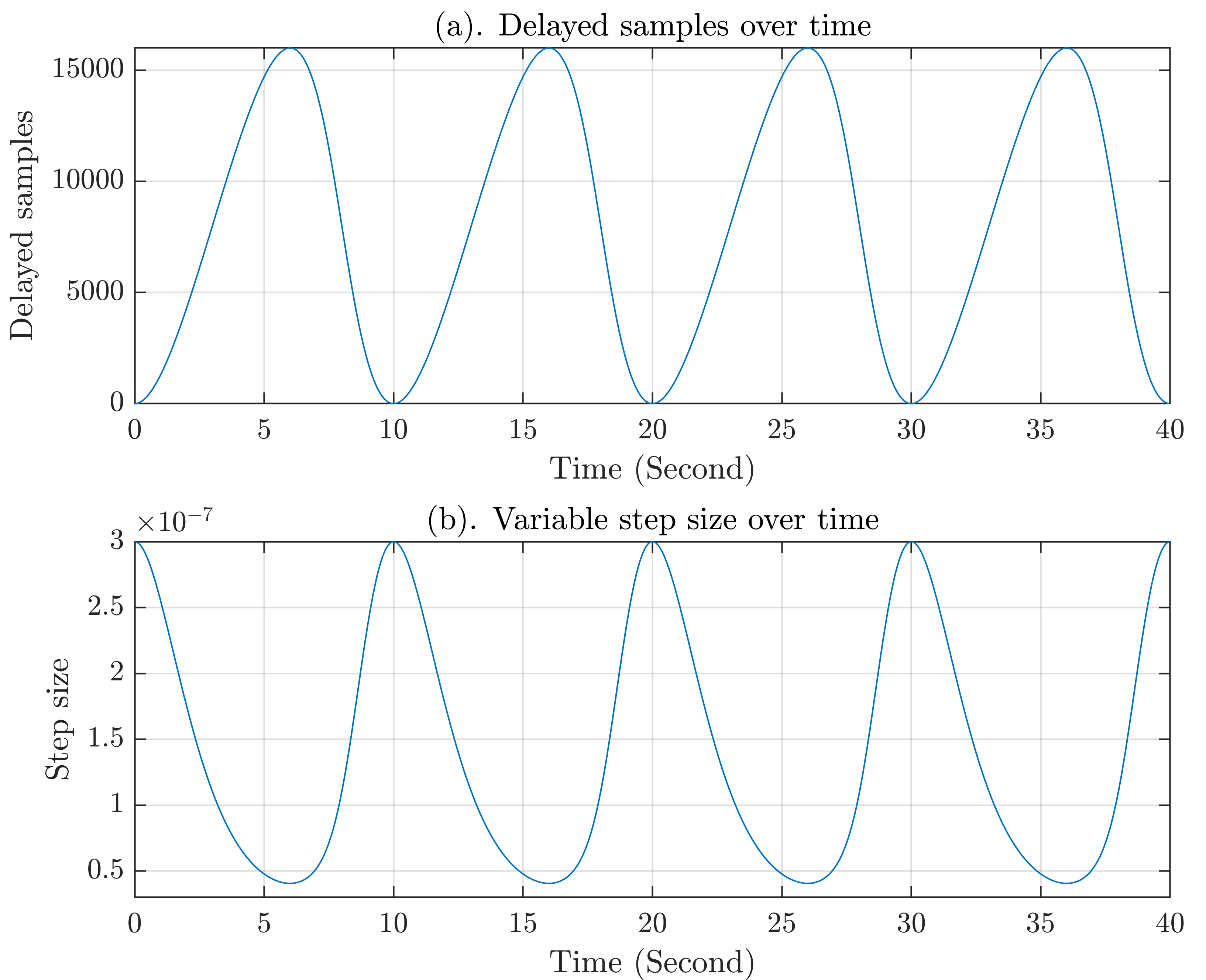}
    \caption{Fluctuating network for all the nodes: (a) communication delay gradually changes over 0-1 seconds; (b) corresponding auto-shrink step size over the varying communication delay.}
    \label{fig:case4_2}
\end{figure}

\begin{figure}[!t]
    \centering
    \includegraphics[width = 0.9\columnwidth, height = 7cm]{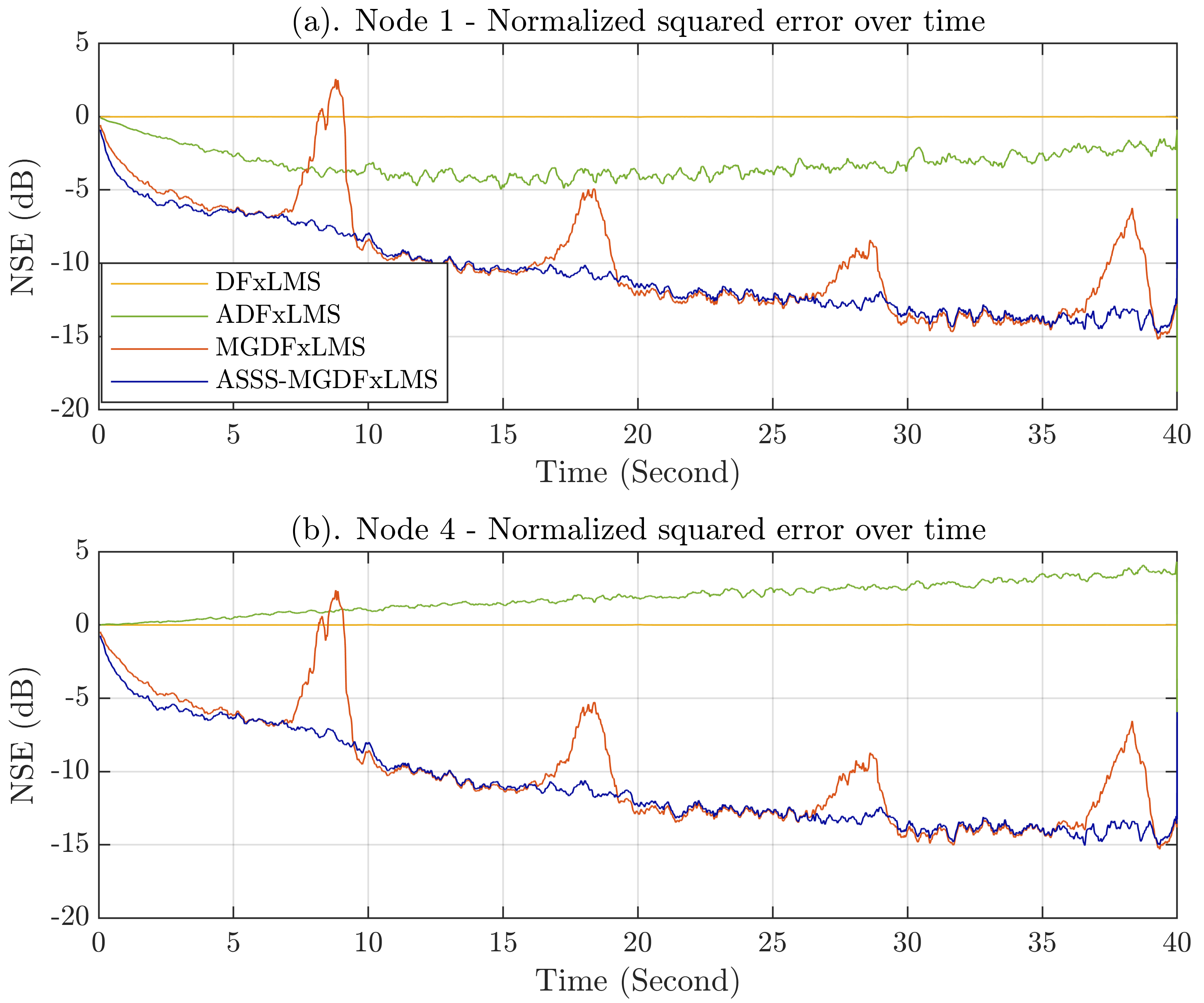}
    \caption{NR performance with different DMCANC algorithms under fluctuating distributed network: (a) NSE over time of node 1; (b) NSE over time of node 4.}
    \label{fig:case4}
\end{figure}

The network latency is subject to fluctuations in practice. Network latency exceeding 500 milliseconds is generally regarded as an extreme network environment, while network latency exceeding 1 second may be perceived as a network failure. In order to further illustrate the efficacy of the ASSS-MGDFxLMS algorithm, we presuppose that the network latency undergoes a gradual transition from $0$ to $1$ seconds, as defined as
\begin{equation}\label{eq:fluctlatency}
    \Delta(n) = \lfloor(\sin{(2\pi\times\frac{0.1n}{f}-\frac{\pi}{2}})+1)\times8000 \rceil,
\end{equation}
where $\lfloor\cdot\rceil$ represents the rounding operation. Fig.~\ref{fig:case4_2} shows the fluctuation of the network and the corresponding variable step size. It can be seen from Fig.~\ref{fig:case4} that no noise reduction occurs in the DFxLMS algorithm, while the ADFxLMS and MGDFxLMS algorithm becomes unstable. The proposed ASSS-MGDFxLMS algorithm can converge and provide a more stable noise reduction performance. Therefore, the proposed algorithm is more robust in coping with fluctuating communication delays.

\begin{figure}[!t]
    \centering
    \includegraphics[width = 0.9\columnwidth,height = 6.5cm]{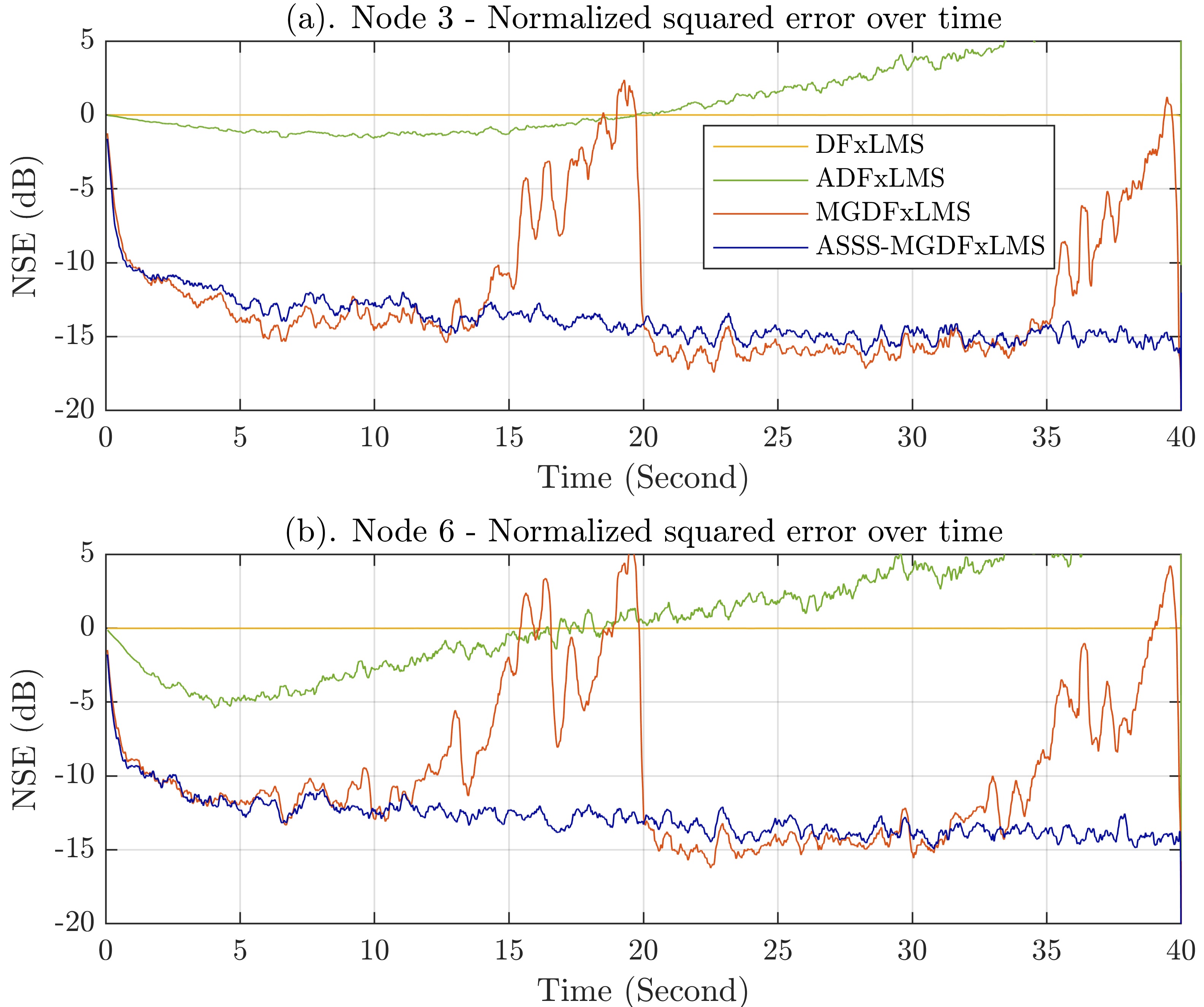}
    \caption{NR performance with different DMCANC algorithms if each node has its individual communicate delay: (a) NSE over time of node 3; (b) NSE over time of node 6.}
    \label{fig:case5}
\end{figure}

\subsection{Noise reduction performance of ANC system when nodes owning different communication delays}
By considering the more general case that each node suffers different delays between the communication, the delayed sample for the $k$th node is defined as
\begin{equation}\label{eq:fluctlatencynode}
    \Delta_k(n) = \lfloor(\sin{(2\pi\times\frac{0.05n}{f}\times k-\frac{\pi}{2}})+1)\times8000 \rceil, 
\end{equation}
where $k$ denotes the index of the node. 

The ASSS-MGDFxLMS was initially configured with a step size of $1\times10^{-6}$, while the other algorithm was configured with a step size of $7\times10^{-7}$. Fig.~\ref{fig:case5} further demonstrates that the proposed ASSS-MGDFxLMS algorithm can achieve a satisfactory noise reduction level despite each node having a distinct communication delay. Conversely, the other algorithms may be adversely affected by unstable distributed networks. Therefore, the proposed ASSS-MGDFxLMS algorithm has the potential to effectively mitigate the system instability that is a result of communication latency.

\section{Conclusion} \label{sec:conclusion}
This paper presented a robust DMCANC algorithm, where compensation filters were applied to make up the difference in the secondary paths between nodes. Instead of sharing local control filters, the local gradients of each node were shared to directly update the global control filters. The proposed algorithm exhibits comparable noise reduction performance to the conventional centralized MCANC algorithm at steady-state.

Furthermore, the theoretical analysis reveals that communication delays across nodes in the actual scenario will decrease the maximum step size and, hence, undermine the stability of the distributed ANC system. To solve this issue, we developed the auto-shrink step size (ASSS) strategy, which can shrink the step size in accordance with the communication delay.     
The simulation results confirmed the efficacy of the proposed MGDFxLMS algorithm compared to previous distributed algorithms. It also demonstrated strong resilience to various communication delays when using the ASSS approach, which is of practical importance.


%



\section*{{Acknowledgment}}
{This research is supported by the Ministry of Education, Singapore, under its Academic Research Fund Tier 2 (MOE-T2EP20221-0014) and (MOE-T2EP50122-0018).}


\ifCLASSOPTIONcaptionsoff
  \newpage
\fi



\bibliographystyle{IEEEtran}
\bibliography{IEEEabrv,ref}
\end{document}